\documentclass[conference]{IEEEtran}
\IEEEoverridecommandlockouts

\usepackage{cite}
\usepackage{amsmath,amssymb,amsfonts}
\usepackage{algorithmic}
\usepackage{graphicx}
\usepackage{textcomp}
\usepackage{xcolor}
\usepackage{colortbl}
\usepackage{booktabs}
\usepackage{xurl}
\usepackage{romannum}
\usepackage{booktabs}
\usepackage{multirow}
\usepackage{etoolbox}
\usepackage[hidelinks]{hyperref}
\usepackage{orcidlink}

\makeatletter
\patchcmd{\@makecaption}
  {\scshape}
  {\upshape}
  {}
  {\PackageWarning{caption-fix}{Caption patch failed}}
\makeatother

\providecommand{\ind}{\mathbb{I}}
\providecommand{\R}{\mathbb{R}}
\providecommand{\Ent}{\mathrm{Ent}}

\providecommand{\Attn}{\mathrm{Attn}}

\providecommand{\sig}{\sigma}
\providecommand{\BCE}{\mathrm{BCE}}
\def\BibTeX{{\rm B\kern-.05em{\sc i\kern-.025em b}\kern-.08em
    T\kern-.1667em\lower.7ex\hbox{E}\kern-.125emX}}

\newcommand{\SOneAuthor}[4]{%
  \begin{minipage}[t]{0.32\textwidth}
    \vspace{0pt}\centering
    \normalfont\fontsize{10}{12}\selectfont
    #1\,\orcidlink{#4}\par
    \vspace{2pt}
    \fontsize{9}{11}\selectfont
    \textit{#2}\\
    \textit{South China University of Technology}\\
    Guangzhou, China\\
    #3\par
  \end{minipage}%
}

\begin{document}
\title{S1Pert: Scalable Target-Specific Gene Response Prediction with System-One Intelligence}

\author{%
\begin{minipage}{\textwidth}
\centering
\SOneAuthor{1\textsuperscript{st} Xiaoqi Sheng$^{\dagger}$}
  {School of Future Technology}
  {xqsheng@scut.edu.cn}
  {0000-0002-2929-5805}%
\SOneAuthor{2\textsuperscript{nd} Hou Wang$^{\dagger}$}
  {School of Computer Science\\and Engineering}
  {cswh@mail.scut.edu.cn}
  {0009-0001-8627-236X}%
\hfill
\SOneAuthor{3\textsuperscript{rd} Yan Liang}
  {School of Future Technology}
  {202464870771@mail.scut.edu.cn}
  {0009-0001-6998-3987}%
\hfill\par\vspace{12pt}
\par
\SOneAuthor{4\textsuperscript{th} Yutong Li}
  {School of Future Technology}
  {202610197035@mail.scut.edu.cn}
  {0009-0006-6764-0776}%
\SOneAuthor{5\textsuperscript{th} Wenhao Du}
  {School of Future Technology}
  {202521063846@mail.scut.edu.cn}
  {0009-0008-0462-0983}%
\hfill
\SOneAuthor{6\textsuperscript{th} Tinghe Zhang$^{*}$}
  {School of Future Technology}
  {zhangth@scut.edu.cn}
  {0000-0002-4595-6167}%
\hfill\par\vspace{12pt}
\par
\hspace*{\fill}
\SOneAuthor{7\textsuperscript{th} Jiawen Liu$^{*}$}
  {School of Computer Science\\and Engineering}
  {202610191544@mail.scut.edu.cn}
  {0009-0002-6827-6049}%
\hspace{0.12\textwidth}
\SOneAuthor{8\textsuperscript{th} Hongmin Cai$^{*}$}
  {School of Future Technology}
  {hmcai@scut.edu.cn}
  {0000-0002-2747-7234}%
\hspace*{\fill}\par
\end{minipage}%
\thanks{$^{\dagger}$ Xiaoqi Sheng and Hou Wang contributed
equally to this work (co-first authors).}%
\thanks{$^{*}$ Corresponding authors:
Tinghe Zhang (zhangth@scut.edu.cn), Jiawen Liu (202610191544@mail.scut.edu.cn) and Hongmin Cai (hmcai@scut.edu.cn)}%
}

\maketitle

\begin{abstract}
Predicting transcriptional responses to unseen genetic perturbations is essential for extending CRISPR screens across vast target--gene combinations. However, models can achieve strong aggregate predictive performance by reproducing historical gene-level response frequencies without adequately distinguishing the effects of different perturbations. Reliable prediction at the transcriptome scale is further challenged by uncertain response labels and redundant computation across millions of target--gene pairs. 
We present S1Pert, a System-One framework for target-specific gene response prediction with non-autoregressive online inference. S1Pert combines directional Welch tests and equivalence testing to distinguish statistically supported responses from inconclusive observations. A frozen System-One encoder extracts reusable semantic representations of perturbation targets, query genes, and historical evidence. Compact pair-specific interactions are then computed and cached offline.
A lightweight hierarchical cross-attention decoder integrates these representations with cross-target pairwise ranking and synchronized historical-evidence dropout to reduce reliance on gene-level response frequencies.
Extensive experiments demonstrate that S1Pert improves target-specific response discrimination while maintaining reliable prediction and efficient inference across diverse cellular contexts. These findings highlight the potential of System-One decision learning for scalable perturbation-response prediction beyond historical gene-level response tendencies. The source code is publicly available at \href{https://github.com/liujiawen-jpg/S1Pert}{https://github.com/liujiawen-jpg/S1Pert}.

\end{abstract}


\section{Introduction}

High-throughput CRISPR screens have enabled systematic measurement of transcriptional responses to genetic perturbations~\cite{dixit2016perturb,replogle2022mapping}, yet the vast number of possible perturbation--gene combinations remains beyond experimental capacity~\cite{roohani2024gears}. Computational approaches seek to predict responses to unseen perturbation targets~\cite{jiang2026learning}, either by estimating continuous expression profiles from gene representations and biological networks~\cite{11603313,roohani2024gears,wenkel2026txpert,zhu2025scouter,cai2025genetic} or by classifying response directions using biological knowledge and language-based reasoning~\cite{wu2025perturbqa,phillips2025synthpert,wang2026aroma,kim2026pertreason,xiang2026aligned}. Despite advances in predictive performance, strong results on aggregate metrics do not necessarily establish whether a model distinguishes the effects of different perturbations on the same gene or merely reproduces its historical response tendency.

Genes differ substantially in how frequently they respond to genetic interventions~\cite{nadig2025transcriptome,replogle2022mapping}. Some genes are repeatedly upregulated or downregulated across diverse perturbations, allowing models to achieve strong aggregate performance by learning historical gene-level response frequencies. Such predictions may be accurate on average while remaining largely insensitive to the identity of the perturbed target. Consistent with this, recent benchmarks have shown that simple baselines can remain competitive with more complex perturbation predictors~\cite{ahlmann2025deep,yuan2026plausibility}. Even under target-held-out evaluation, a gene's historical response tendency remains informative across unseen interventions, so aggregate metrics alone do not establish whether predictions depend on the perturbed target. Reliable unseen-target prediction therefore requires distinguishing how the same query gene responds to different perturbations, both in whether its expression changes and in the direction of that change.

\begin{figure}[!t]
    \centering
    \includegraphics[width=\columnwidth]{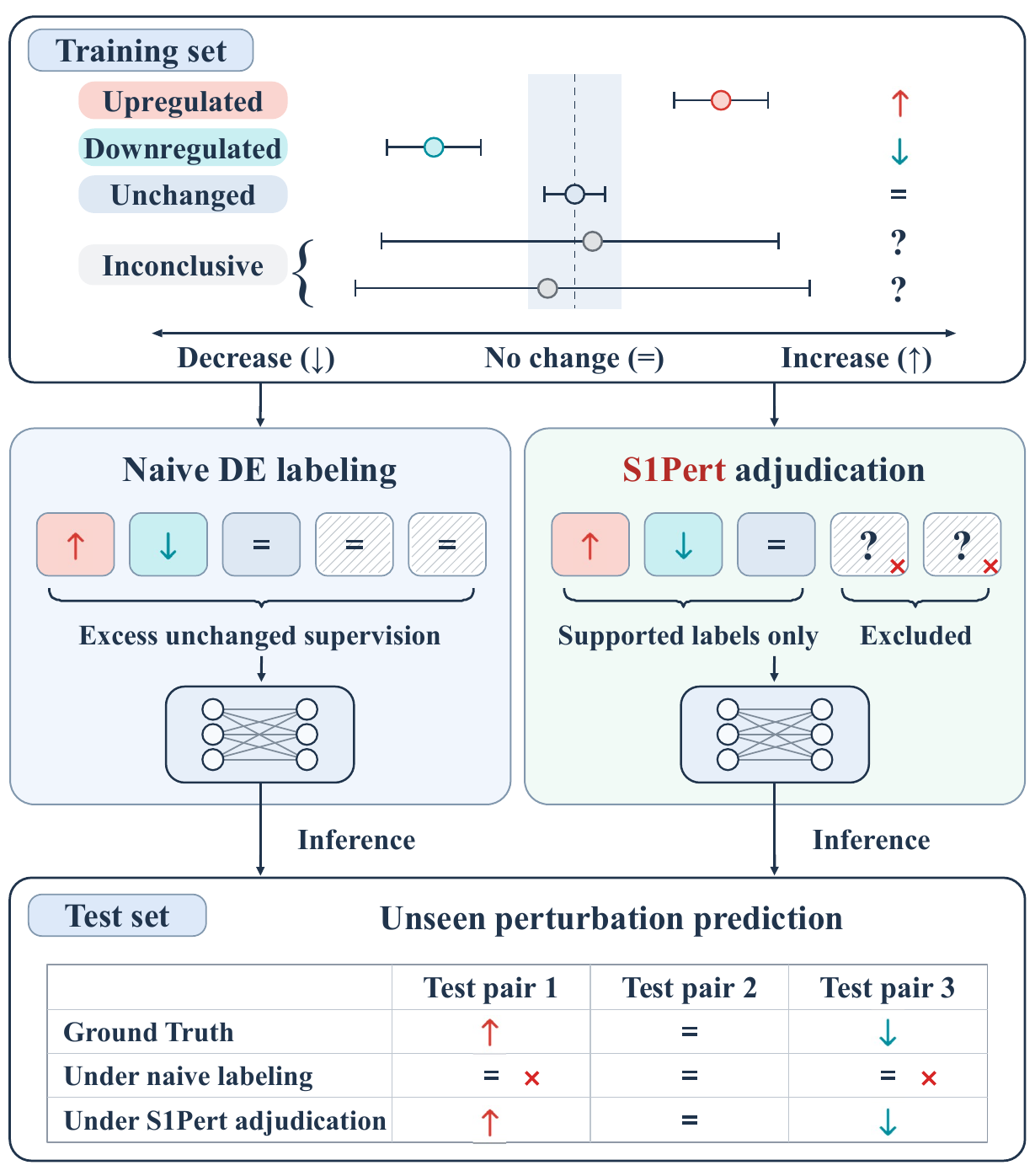}
    \caption{Label adjudication for unseen-perturbation prediction. Naive DE labeling treats inconclusive responses as unchanged, biasing predictions toward no change. S1Pert excludes inconclusive training pairs and retains statistically supported labels. Observations and predictions are schematic.}
    \label{fig:label-motivation}
\end{figure}

Reliable target-specific prediction at transcriptome scale also faces two practical challenges: label uncertainty and redundant computation. Categorical perturbation-response datasets may classify statistically nonsignificant results as unchanged~\cite{wu2025perturbqa}, although nonsignificance does not establish equivalence~\cite{barry2024robust,lakens2017equivalence}. As illustrated in Fig.~\ref{fig:label-motivation}, observations with wide uncertainty intervals straddle the no-change equivalence zone and remain statistically inconclusive. Naive differential expression (DE) labeling merges these inconclusive responses into the unchanged class, introducing excess unchanged supervision~\cite{smirnova2022nessy,yang2023learning,li2021cleanml} and distorting historical response priors, which biases the trained predictor toward outputting no change for genuinely upregulated or downregulated test pairs. At scale, millions of target--gene pairs share gene annotations, perturbation descriptions, and historical evidence templates. Repeatedly encoding shared text or generating autoregressive rationales for individual pairs incurs substantial computational costs~\cite{chen2026bond,du2026c2kv,phillips2025synthpert,kim2026progressive}.

Here we introduce S1Pert, a System-One perturbation-response predictor that combines statistically adjudicated supervision, reusable semantic representations, and single-pass, non-autoregressive decision-making. To improve label reliability, S1Pert integrates directional Welch tests with two-one-sided equivalence testing (TOST)~\cite{schuirmann1987comparison,lakens2017equivalence}, retaining supported upregulation ($\uparrow$), equivalence-supported unchanged responses ($=$), and supported downregulation ($\downarrow$) while excluding statistically inconclusive observations ($?$) from training supervision (Fig.~\ref{fig:label-motivation}). S1Pert represents perturbation targets and readout genes through separate Condition and Query streams. A Historical Evidence stream summarizes out-of-fold response tendencies across training perturbations. To reduce redundant encoding, a frozen Jev model~\cite{mapika2026decider4b} processes $|\mathcal{T}|+|\mathcal{G}|+109$ unique full-length inputs for $|\mathcal{T}|$ perturbation targets, $|\mathcal{G}|$ query genes, and 109 historical-evidence templates. Pair-specific interactions are preserved through compact 36-token sequences processed by the encoder's upper eight layers, with both evidence-present and evidence-masked representations cached offline. During online inference, a lightweight hierarchical cross-attention decoder estimates expression-change and conditional direction probabilities without autoregressive generation. Cross-target pairwise ranking encourages the model to distinguish different perturbations acting on the same query gene, while synchronized historical-evidence dropout reduces reliance on historical response frequencies.

Across four cellular benchmarks, S1Pert consistently outperforms state-of-the-art baselines in directional decision accuracy, within-gene target discrimination, and online inference efficiency. The main contributions of this work are summarized as follows:
\begin{itemize}
    \item \textbf{Reliable response supervision.}
    We develop an equivalence-adjudicated labeling protocol that distinguishes statistically supported unchanged responses from inconclusive observations, with label fidelity evaluated using disjoint cell pools.

    \item \textbf{Scalable multi-stream decision architecture.}
    We introduce a cached System-One framework that reduces full-prompt encoding from quadratic target--gene pairs to linear single-stream inputs while preserving cross-stream interactions for fast, non-autoregressive hierarchical decoding.

    \item \textbf{Target-specific response learning.}
     We integrate cross-target pairwise ranking with synchronized historical-evidence dropout to suppress gene-frequency shortcuts, substantially improving within-gene polarity discrimination and robustness under conflicting or missing priors.
\end{itemize}

\section{Related Work}

\subsection{Target-Specific Perturbation Prediction}
 computational frameworks for unseen-target perturbation prediction broadly follow two paradigms~\cite{rood2024toward,peidli2024scperturb}.
Continuous profile predictors such as GEARS~\cite{roohani2024gears} and STATE~\cite{State2026} regress post-perturbation transcriptomic states from gene co-expression graphs, biological ontologies, and context-aware state transitions~\cite{cai2025genetic}.
Complementarily, discrete response classifiers such as ALIGNED~\cite{xiang2026aligned} integrate symbolic regulatory rules or multimodal priors~\cite{wang2026aroma} to categorize gene-level outcomes into upregulation, invariance, or downregulation.

However, high aggregate accuracy under target-held-out evaluation does not guarantee genuine target-level discrimination~\cite{yuan2026plausibility}.
In data mining and recommender systems, models trained with pointwise supervision over skewed relational interactions are known to suffer from popularity and frequency bias, achieving low aggregate error by memorizing entity-level marginal frequencies rather than resolving pair-specific interactions~\cite{wei2021model,tan2025taming}.
An analogous shortcut arises in perturbation screening, where predictors can minimize overall loss by reciting a query gene's dominant historical response tendency across training interventions while remaining insensitive to the perturbed target.
Extending counterfactual and deconfounded ranking principles from the data engineering literature~\cite{wei2021model,tan2025taming} to biological screen modeling, S1Pert decouples target, query, and prior streams and jointly optimizes cross-target pairwise ranking with synchronized evidence dropout to recover target-specific regulatory polarity.

\subsection{Statistical Data Cleaning and Materialized LLM Inference}
Downstream model reliability is fundamentally bounded by supervision quality, motivating extensive data engineering research on data cleaning for statistical modeling~\cite{krishnan2016activeclean}, probabilistic data repair~\cite{rekatsinas2017holoclean}, and learning under noisy labels~\cite{yang2023learning}.
In single-cell perturbation screens, standard differential expression pipelines~\cite{love2014moderated,barry2024robust,heumos2026pertpy} default to an unchanged label whenever directional hypotheses fail to reach significance~\cite{wang2026aroma,wei2026vcworld}, conflating low-depth sampling noise with biological invariance.
To separate genuine invariance from insufficient statistical power, S1Pert adapts the two-one-sided tests (TOST) framework~\cite{schuirmann1987comparison} into a pre-training data adjudication operator, combining scaled Welch statistics~\cite{welch1947generalization,satterthwaite1946approximate} with equivalence bounds to filter inconclusive observations before supervision and prior construction.

Integrating pretrained foundation models into perturbation screening introduces a second data engineering bottleneck: scalable execution over millions of entity pairs.
While scGPT~\cite{cui2024scgpt} pretrains single-cell transformers, GenePT~\cite{chen2025genept,chen2024geneptembeddings} embeds functional text descriptions, and SUMMER~\cite{wu2025perturbqa} prompts language models with biological knowledge and empirical priors, applying large language models across Cartesian target--gene spaces incurs severe redundancy from repeatedly encoding shared context and decoding autoregressive rationales~\cite{yang2025earn}.
Recent database and systems studies address bulk LLM analytics through declarative semantic operators~\cite{patel2025lotus,liu2025palimpzest}, flash-resident KV-cache management~\cite{shin2026matkv}, and modular attention reuse across overlapping prompts~\cite{gim2024prompt}.
Building on these view-materialization and state-reuse principles, S1Pert adopts a System-One decision paradigm~\cite{mapika2026decider4b} that factorizes pair-level inference into linear single-stream landmark precomputation and compact upper-layer joint materialization, enabling fast, non-autoregressive hierarchical decoding over transcriptome-scale target--gene combinations.

\section{Methodology}

\begin{table}[!t]
\centering
\begingroup
\fontsize{9}{11}\selectfont
\setlength{\tabcolsep}{0pt}
\renewcommand{\arraystretch}{1.05}
\setlength{\lineskip}{2pt}
\setlength{\lineskiplimit}{1pt}
\algsetup{indent=1em,linenosize=\scriptsize}
\providecommand{\ind}{\mathbb{I}}
\begin{tabular}{@{}p{\columnwidth}@{}}
\toprule
\textbf{Algorithm 1} Statistical label adjudication for target $i$ across $\mathcal{G}_\ell$ \\
\midrule
\raggedright
\textbf{Input:}
Raw UMI matrix $\boldsymbol{X}$ and cell total UMI counts
$\{U_a\}$; control and target-$i$ cell sets
$\mathcal{C}_c,\mathcal{C}_i$;
query gene set $\mathcal{G}_\ell$;
equivalence margin $\rho=1.2$ and FDR threshold $\alpha=0.05$.
Normalized expression is $z_{rj}=10^4X_{rj}/U_r$.\par
\smallskip
\textbf{Output:} Adjudicated labels
$\boldsymbol{y}_i=\{y_{ij}\}_{j\in\mathcal{G}_\ell}$,
with $y_{ij}\in\{0,1,2,\bot\}$.
\begin{algorithmic}[1]
\raggedright
\STATE \textbf{Step 1:} Scaled Welch and TOST $p$-values
\FOR{each $j\in\mathcal{G}_\ell$}
    \FOR{each $g\in\{c,i\}$}
        \STATE
        $\mu_{jg} =\frac{1}{n_g}\sum_{r\in\mathcal{C}_g}z_{rj},
        \quad
        m_{jg}=\sum_{r\in\mathcal{C}_g}\ind[X_{rj}>0]$
        \STATE
        $s_{jg}^2 =\frac{1}{n_g-1} \sum_{r\in\mathcal{C}_g}(z_{rj}-\mu_{jg})^2,
        \quad
        e_{jg}=s_{jg}^2/n_g$
    \ENDFOR
    \IF{$j\neq i\land n_c,n_i\ge20\land m_{jc},m_{ji}\ge10$\\
        \hspace{\algorithmicindent}$\land\mu_{jc},\mu_{ji},e_{jc},e_{ji}>0$}
        \FOR{each $a\in\{\rho,1/\rho\}$}
            \STATE $\Delta_{ij}(a)=\mu_{ji}-a\mu_{jc},\quad
            V_{ij}^2(a)=e_{ji}+a^2e_{jc}$
            \STATE $T_{ij}(a)=\Delta_{ij}(a)/V_{ij}(a)$
            \STATE $\displaystyle\nu_{ij}(a)=
            \frac{V_{ij}^4(a)}{e_{ji}^2/(n_i-1)+a^4e_{jc}^2/(n_c-1)}$
        \ENDFOR
        \STATE $p_{ij}^{\mathrm{up}}=
        1-F_{\nu_{ij}(\rho)}\bigl(T_{ij}(\rho)\bigr)$
        \STATE $p_{ij}^{\mathrm{down}}=
        F_{\nu_{ij}(1/\rho)}\bigl(T_{ij}(1/\rho)\bigr)$
        \STATE $p_{ij}^{\mathrm{eq}}=
        \max\bigl\{1-p_{ij}^{\mathrm{down}},\,1-p_{ij}^{\mathrm{up}}\bigr\}$
    \ELSE
        \STATE $p_{ij}^{\mathrm{up}}=p_{ij}^{\mathrm{down}}=p_{ij}^{\mathrm{eq}}=1$
    \ENDIF
\ENDFOR
\smallskip
\STATE \textbf{Step 2:} Benjamini--Hochberg FDR control\\
       and label assignment
\STATE $\{q_{ij}^{\mathrm{up}},q_{ij}^{\mathrm{down}},q_{ij}^{\mathrm{eq}}\}_{j\in\mathcal{G}_\ell}
       =\mathrm{BH}\bigl(\{p_{ij}^{\mathrm{up}},p_{ij}^{\mathrm{down}},p_{ij}^{\mathrm{eq}}\}_{j\in\mathcal{G}_\ell}\bigr)$
\FOR{each $j\in\mathcal{G}_\ell$}
    \IF{$q_{ij}^{\mathrm{up}}\le\alpha\land\Delta_{ij}(\rho)>0$}
        \STATE $y_{ij}=0$\quad (Upregulated)
    \ELSIF{$q_{ij}^{\mathrm{eq}}\le\alpha\land\Delta_{ij}(1/\rho)>0\land\Delta_{ij}(\rho)<0$}
        \STATE $y_{ij}=1$\quad (Unchanged)
    \ELSIF{$q_{ij}^{\mathrm{down}}\le\alpha\land\Delta_{ij}(1/\rho)<0$}
        \STATE $y_{ij}=2$\quad (Downregulated)
    \ELSE
        \STATE $y_{ij}=\bot$\quad (Inconclusive)
    \ENDIF
\ENDFOR
\RETURN $\boldsymbol{y}_i$
\end{algorithmic}\tabularnewline
\bottomrule
\end{tabular}
\endgroup
\end{table}

\subsection{Problem Formulation}
For each cellular context $\ell$, let $\mathcal{T}_\ell$ denote the set of perturbation targets and $\mathcal{G}_\ell$ the set of measured query genes, where $G = |\mathcal{G}_\ell|$. The targets are partitioned into mutually exclusive training, validation, and test sets. 
For each target $i\in\mathcal{T}_\ell$ and query gene $j\in\mathcal{G}_\ell$, the predictor returns a three-class probability vector $f_\theta(i,j,\ell) = \bigl[\hat p_{ij,0},\hat p_{ij,1},\hat p_{ij,2}\bigr]^{\mathsf T} \in \Delta^2$, corresponding to upregulation ($y_{ij}=0$), verified invariance ($y_{ij}=1$), or downregulation ($y_{ij}=2$) relative to untreated controls. Test-target perturbed readouts are withheld from the predictor, which instead uses untreated-control statistics, curated gene-function annotations, and response propensities derived from the training split.

The predictive performance of $f_\theta$ depends on the reliability of its reference labels $y_{ij}$. In standard differential-expression analyses, a gene is often labeled unchanged when neither directional hypothesis reaches statistical significance~\cite{wang2026aroma,wei2026vcworld}. Such a rule treats a lack of significant evidence for change as evidence of biological invariance, although the result may instead reflect finite-cell sampling noise~\cite{lakens2017equivalence}. S1Pert addresses this ambiguity with the two-stage adjudication protocol in \textbf{Algorithm~1}. In the protocol, $X_{aj}$ and $U_a$ denote the raw UMI count of gene $j$ and the total UMI count in cell $a$, respectively. The sets $\mathcal{C}_c$ and $\mathcal{C}_i$ comprise untreated control cells and cells receiving target $i$, with sample sizes $n_c$ and $n_i$, respectively.
\textbf{Step~1} estimates the sample mean $\mu_{jg}$, nonzero count $m_{jg}$, unbiased variance $s_{jg}^2$, and sampling variance of the mean $e_{jg}=s_{jg}^2/n_g$ for each group $g\in\{c,i\}$. Gene--target pairs that fail the minimum support criteria receive unit $p$-values for all three tests. S1Pert then evaluates the multiplicative equivalence boundaries $a\in\{\rho,1/\rho\}$ with $\rho=1.2$. For each boundary $a$, the estimated variance $V_{ij}^2(a)$ and standard error $V_{ij}(a)=\sqrt{V_{ij}^2(a)}$ standardize the scaled mean difference $\Delta_{ij}(a)$. The resulting Welch statistic $T_{ij}(a)$~\cite{welch1947generalization} uses the Welch--Satterthwaite degrees of freedom $\nu_{ij}(a)$~\cite{satterthwaite1946approximate}, with probabilities evaluated through the Student-$t$ CDF $F_\nu(\cdot)$.
The directional tests quantify evidence for responses above $\rho\mu_{jc}$ or below $\mu_{jc}/\rho$, producing $p_{ij}^{\mathrm{up}}$ and $p_{ij}^{\mathrm{down}}$, respectively. In parallel, the two one-sided tests procedure evaluates equivalence within $[1/\rho,\rho]\mu_{jc}$ and produces $p_{ij}^{\mathrm{eq}}$~\cite{schuirmann1987comparison}. This design distinguishes affirmative evidence of invariance from directional nonsignificance.
\textbf{Step~2} applies Benjamini--Hochberg FDR control at $\alpha=0.05$ to the $3|\mathcal{G}_\ell|$ hypotheses for each target $i$. FDR-supported outcomes are labeled as upregulation $y_{ij}=0$, verified invariance $y_{ij}=1$, or downregulation $y_{ij}=2$. Remaining pairs receive the inconclusive label $y_{ij}=\bot$. Their exclusion from training and primary evaluation limits the propagation of uncertainty caused by UMI dropout and heterogeneous guide capture into historical-response priors and target prioritization~\cite{barry2024robust,nadig2025transcriptome}.

\subsection{Overview}
\begin{figure*}[!t]
    \centering
    \includegraphics[width=\textwidth]{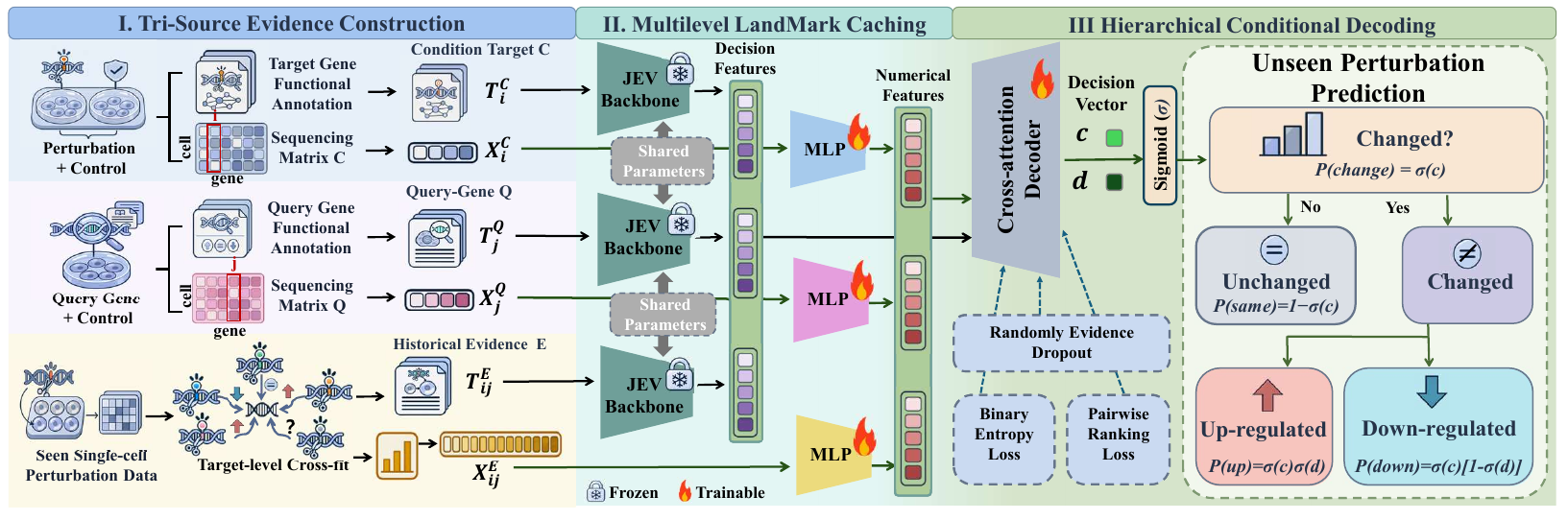}
    \caption{Overview of the S1Pert framework. (I) \textbf{Tri-Source Evidence Construction} pairs gene functional annotations and control sequencing matrices with target-level cross-fitted historical statistics to construct textual prompts $(T_i^C, T_j^Q, T_{ij}^E)$ and continuous covariates $(\boldsymbol{x}_i^C, \boldsymbol{x}_j^Q, \boldsymbol{x}_{ij}^E)$ for the condition target ($C$), query gene ($Q$), and historical evidence ($E$). (II) \textbf{Multilevel Landmark Caching} extracts and caches compressed semantic decision features offline via a parameter-shared frozen Jev backbone, while projecting continuous covariates into numerical features through trainable stream-specific MLPs. (III) \textbf{Hierarchical Conditional Decoding} integrates the decision and numerical features through a cross-attention decoder—trained with random evidence dropout, binary cross-entropy loss, and pairwise ranking loss—to output a decision vector $(c, d)$ for hierarchical unseen perturbation prediction.}
    \label{fig:framework}
\end{figure*}

Fig.~\ref{fig:framework} illustrates the overall framework of the proposed S1Pert, which consists of three key stages: Tri-Source Evidence Construction (Fig.~\ref{fig:framework}-\Romannum{1}), Multilevel Landmark Caching (Fig.~\ref{fig:framework}-\Romannum{2}), and Hierarchical Conditional Decoding (Fig.~\ref{fig:framework}-\Romannum{3}).
Tri-Source Evidence Construction organizes functional annotations, control sequencing matrices, and target-level cross-fitted statistics into textual prompts $(T_i^C, T_j^Q, T_{ij}^E)$ and continuous covariates $(\boldsymbol{x}_i^C, \boldsymbol{x}_j^Q, \boldsymbol{x}_{ij}^E)$.
Multilevel Landmark Caching extracts compressed decision features offline via a parameter-shared frozen Jev backbone while projecting the continuous covariates into numerical features through trainable MLPs.
Conditioned on both feature streams, Hierarchical Conditional Decoding uses a cross-attention decoder to predict a decision vector $(c, d)$ that factorizes expression-change probability $\sigma(c)$ and directional polarity $\sigma(d)$ for unseen perturbation prediction.
Finally, random evidence dropout, binary cross-entropy loss, and pairwise ranking loss jointly optimize the decoder to preserve target-specific separability and suppress gene-frequency shortcuts.

\begin{table}[!t]
\centering
\caption{Structured prompts and examples for the tri-source streams in S1Pert.}
\label{tab:tri_source_prompts}
\begingroup
\fontsize{8.5}{10}\selectfont
\setlength{\tabcolsep}{5pt}
\renewcommand{\arraystretch}{1.08}
\begin{tabular}{>{\raggedright\arraybackslash}p{\dimexpr\columnwidth-2\tabcolsep\relax}}
\toprule
\rowcolor{black!8}
\textbf{Shared Task Prefix ($\mathcal{P}_{\mathrm{task}}$)} \\
Prepended verbatim to both $T_i^C$ and $T_j^Q$.\par
\textbf{Task Definition:} Predict population-average RNA direction
(upregulated / unchanged / downregulated vs.\ $10^4$-UMI controls)
after CRISPRi; missing knowledge $\nRightarrow$ unchanged. \\

\midrule
\rowcolor{black!8}
\textbf{Condition Stream Prompt ($T_i^C$)} \\
\textbf{Context:} $\mathcal{P}_{\mathrm{task}}$ + Cell context:
\texttt{\{cell\}}. CRISPRi targets \texttt{\{sym\}}
(\texttt{\{id\}}); intended action is reduced target transcription. \\
\addlinespace[2pt]
\textbf{Target Control \& Function:}
$\text{mean}=\texttt{\{}\mu_{\iota(i)c}\texttt{\}}$,
$\text{var}=\texttt{\{}s_{\iota(i)c}^2\texttt{\}}$,
$\text{det}=\texttt{\{}\delta_{\iota(i)}^c\texttt{\}}$;
\texttt{\{target\_function\}}. \\
\addlinespace[2pt]
\textbf{Example:} \texttt{rpe1} $|$
\texttt{RPL11 (ENSG00000142676)} $|$
\texttt{18.4, 6.82, 0.96} $|$
\textit{60S ribosomal subunit L11; regulates MDM2--p53 signaling.} \\

\midrule
\rowcolor{black!8}
\textbf{Query Stream Prompt ($T_j^Q$)} \\
\textbf{Query State \& Function:}
$\mathcal{P}_{\mathrm{task}}$ + Query gene \texttt{\{sym\}}
(\texttt{\{id\}}) control
$\text{mean}=\texttt{\{}\mu_{jc}\texttt{\}}$,
$\text{var}=\texttt{\{}s_{jc}^2\texttt{\}}$,
$\text{det}=\texttt{\{}\delta_j^c\texttt{\}}$;
\texttt{\{query\_function\}}. \\
\addlinespace[2pt]
\textbf{Question \& Options:} Population mean RNA change vs.\ controls
$\circ$~\textbf{Upregulated:} $>1.2$;
$\circ$~\textbf{Unchanged:} $[1/1.2,1.2]$;
$\circ$~\textbf{Downregulated:} $<1/1.2$. \\
\addlinespace[2pt]
\textbf{Example:} \texttt{CDKN1A (ENSG00000124762)} $|$
\texttt{2.31, 3.85, 0.74} $|$
\textit{Cyclin-dependent kinase inhibitor 1; mediates p53-dependent arrest.} \\

\midrule
\rowcolor{black!8}
\textbf{Historical Evidence Card ($T_{ij}^E$)} \\
$108+1$ discrete templates.\par
\textbf{Evidence Schema:} Across other training targets in
\texttt{\{cell\}}, this query gene \texttt{\{tendency\}} with
\texttt{\{concentration\}} concentration, \texttt{\{support\}} support,
and \texttt{\{source\}} estimate (target-specific mechanisms may deviate). \\
\addlinespace[2pt]
\textbf{Example:} \textit{more often increased} $|$
\texttt{strong} $|$ \texttt{broad} ($\ge\!50$) $|$
\texttt{cell-query specific}. \\
\bottomrule
\end{tabular}
\endgroup
\end{table}

\subsection{Tri-Source Evidence Construction}
Response prediction for an untested CRISPRi target $i$ and a readout gene $j$ depends on the target’s functional context, the query gene’s regulatory and basal-expression properties, and response patterns observed across prior perturbations. Specifically, the basal expression and functional context of target gene $i$ shape the perturbation effect~\cite{replogle2022mapping}. For query gene $j$, regulatory connections inform the response, while basal expression and variability influence its statistical detectability~\cite{dixit2016perturb,barry2024robust}. Historical perturbation profiles further characterize gene-specific responsiveness and stress-response programs shared across interventions~\cite{nadig2025transcriptome,replogle2022mapping}. To represent these complementary factors, S1Pert encodes each target--gene pair $(i,j)$ using condition ($C$), query ($Q$), and historical-evidence ($E$) streams. Each stream integrates its structured prompt schema in Table~\ref{tab:tri_source_prompts} with the associated continuous covariates.

To represent the first two biological factors, S1Pert encodes the upstream perturbation target and downstream readout gene in the condition and query streams, respectively. Each stream pairs a textual prompt ($T_i^C$ or $T_j^Q$) with basal control covariates ($\boldsymbol{x}_i^C$ or $\boldsymbol{x}_j^Q$).
The \textbf{Shared Task Prefix} $\mathcal{P}_{\mathrm{task}}$, shown in the top block of Table~\ref{tab:tri_source_prompts}, defines the three-class directional prediction task using expression counts normalized to $10^4$ UMIs per cell. It also specifies that missing functional annotations do not constitute evidence of unchanged expression. S1Pert prepends this prefix verbatim to both $T_i^C$ and $T_j^Q$, providing consistent task instructions and evidence constraints before cross-stream interaction. This instruction reduces the risk that poorly annotated genes are implicitly represented as transcriptionally invariant.
The second block of Table~\ref{tab:tri_source_prompts} organizes $T_i^C$ into \textbf{Context} and \textbf{Target Control \& Function}. In the RPE1 example, the prompt characterizes RPL11 knockdown by the target's basal profile (\texttt{mean=18.4, var=6.82, det=0.96}) and curated roles in 60S ribosomal assembly and MDM2--p53 signaling. The third block organizes $T_j^Q$ into \textbf{Query State \& Function} and \textbf{Question \& Options}. For CDKN1A, it combines the baseline profile (\texttt{mean=2.31, var=3.85, det=0.74}) and p53-mediated arrest function with the fold-change intervals specified in \textbf{Algorithm~1}, namely $>1.2$, $[1/1.2,1.2]$, and $<1/1.2$.
The prompts encode control summaries as tokens for semantic processing, while tokenization discretizes their continuous magnitudes. Moreover, a target outside the measured gene set $\mathcal{G}_\ell$, indicated by $\iota(i)<0$, lacks a target-indexed representation. S1Pert therefore pairs textual prompts with continuous covariates to retain quantitative information and avoid dependence on a measured-gene index.
S1Pert pairs the prompts with four-dimensional continuous covariates $\boldsymbol{x}_j^Q, \boldsymbol{x}_i^C \in \R^4$ using control detection rate $\delta_j^c = m_{jc}/n_c$ via
\begin{equation}
    \boldsymbol{x}_j^Q =
    \big[\log(1+\mu_{jc}),\, \log(1+s_{jc}^2),\,
    \delta_j^c,\, 1\big]^{\mathsf{T}}.
    \label{eq:numeric-features}
\end{equation}
The condition covariate is defined as
$\boldsymbol{x}_i^C=\boldsymbol{x}_{\iota(i)}^Q$ if
$\iota(i)\ge0$, and $\boldsymbol{x}_i^C=\mathbf{0}_4$
otherwise, where $\iota(i)$ indexes target gene $i$
within $\mathcal{G}_\ell$ and is negative if the target
is unmeasured in the query panel.

Historical response counts provide a training-derived estimate of each readout gene’s response propensity. To maintain a consistent information boundary, these counts are constructed exclusively from $\mathcal{T}_{\ell,\mathrm{train}}$, with validation and test outcomes withheld throughout feature construction. A full-training aggregate would contain the current training target’s label $y_{ij}$, creating a self-inclusion shortcut that would be unavailable for unseen targets.
S1Pert uses five-fold target-level cross-fitting to align training features with the validation and test setting. A split-locked random permutation assigns training targets to five mutually exclusive, approximately equal-sized folds $\{\mathcal{F}_1,\ldots,\mathcal{F}_5\}$. For gene $j$ and label $k\in\{0,1,2\}$, let $C_j^{\mathrm{tr},k}$ be the count across $\mathcal{T}_{\ell,\mathrm{train}}$ and $C_j^{f,k}$ the count within $\mathcal{F}_f$. A training instance $i\in\mathcal{F}_f$ receives the leave-fold-out count $C_{ij}^k = C_j^{\mathrm{tr},k} - C_j^{f,k}$, based on the other 80\% of training targets. For validation and test instances, the feature uses the complete training aggregate $C_{ij}^k = C_j^{\mathrm{tr},k}$. In both cases, query support is computed as $S_{ij}^q = \sum_{k=0}^2 C_{ij}^k$.

In single-cell screens, low transcript abundance and UMI dropout can leave a target--gene pair with no adjudicated out-of-fold observations ($S_{ij}^q = 0$) or only sparse support ($1 \leq S_{ij}^q < 10$). For query cold starts, S1Pert applies a three-tier provenance hierarchy indexed by $o_{ij} \in \{0,1,2\}$. Query-specific counts $C_{ij}^k$ are used whenever $S_{ij}^q > 0$, corresponding to $o_{ij}=0$. When query-specific support is absent, cell-line-wide out-of-fold class totals $B_{ij}^k$ are used if available, corresponding to $o_{ij}=1$. If cell-line-wide support is also absent, the hierarchy backs off to cross-cell-line global class totals $G_{ij}^k$, corresponding to $o_{ij}=2$. These sources define the effective counts $\widetilde{C}_{ij}^k$ and effective support $S_{ij}^{\mathrm{ev}} = \sum_{k=0}^2 \widetilde{C}_{ij}^k$.
Sparse-support queries retain their query-specific counts and provenance level $o_{ij}=0$. Laplace smoothing stabilizes their class-probability estimates by adding one pseudo-count per class:
\begin{equation}
    p_{ij,k}^{\mathrm{ev}} = \frac{\widetilde{C}_{ij}^k + 1}{S_{ij}^{\mathrm{ev}} + 3}, \qquad k \in \{0, 1, 2\},
    \label{eq:evidence-probability}
\end{equation}
The evidence representation also encodes support and uncertainty through normalized log-support $\log(1 + S_{ij}^{\mathrm{ev}}) / D_S$, where $D_S = \max_{\mathrm{train}} \log(1 + S^{\mathrm{ev}})$, normalized Shannon entropy $\Ent(\boldsymbol{p}_{ij}^{\mathrm{ev}}) / \log 3$, and the sorted class propensities $p_{(3)} \geq p_{(2)} \geq p_{(1)}$. These features form the 12-dimensional representation
\begin{equation}
\begin{aligned}
\boldsymbol{x}_{ij}^E = \Big[
& \big(\boldsymbol{p}_{ij}^{\mathrm{ev}}\big)^{\mathsf{T}},\, \frac{\log(1+S_{ij}^{\mathrm{ev}})}{D_S},\, \Ent(\boldsymbol{p}_{ij}^{\mathrm{ev}}),\, p_{(3)},\, p_{(3)}-p_{(2)}, \\
& \big\{\ind[o_{ij}=r]\big\}_{r=0}^2,\, p_{ij,0}^{\mathrm{ev}}-p_{ij,2}^{\mathrm{ev}},\, \ind[S_{ij}^{\mathrm{ev}}>0]\Big]^{\mathsf{T}}.
\end{aligned}
\label{eq:evidence-vector}
\end{equation}
Quantizing $\boldsymbol{p}_{ij}^{\mathrm{ev}}$ and $S_{ij}^{\mathrm{ev}}$ populates the \textbf{Historical Evidence Card} $T_{ij}^E$ in the bottom block of Table~\ref{tab:tri_source_prompts} across four discrete slots: \texttt{\{tendency\}} (4 states at margin $0.08$), \texttt{\{concentration\}} (3 levels: \texttt{weak}, \texttt{moderate}, \texttt{strong} at thresholds $0.50/0.75$ and margin $0.10$), \texttt{\{support\}} (3 strata: \texttt{limited} $<10$ for sparse support, \texttt{moderate} $<50$, \texttt{broad} $\ge 50$), and \texttt{\{source\}} (3 tiers via $o_{ij} \in \{0,1,2\}$), defining a Cartesian schema of $4\times 3\times 3\times 3 = 108$ structured templates alongside an unavailable-evidence card for $S_{ij}^{\mathrm{ev}} = 0$.
Within each cell line, at most 31 templates are reachable. Margin rules exclude \texttt{moderate}/\texttt{strong} concentration for \texttt{mixed} tendencies, while Laplace smoothing precludes \texttt{strong} concentration at very low support. Cell-line fallback $B_{ij}^k$ has \texttt{broad} support ($\ge 50$), and $o_{ij}=2$ is reserved for cross-atlas cold start.

Overall, tri-source construction represents each target--gene pair $(i,j)$ by three structured prompts $(T_i^C, T_j^Q, T_{ij}^E)$ paired with three continuous feature vectors $(\boldsymbol{x}_i^C, \boldsymbol{x}_j^Q, \boldsymbol{x}_{ij}^E)$ for downstream semantic caching and conditional decoding.

\subsection{Multilevel Landmark Caching}

Encoding $(T_i^C, T_j^Q, T_{ij}^E)$ across millions of target--gene pairs requires capturing cross-stream interactions without running quadratic full-text encoding over all $(i,j)$ pairs.
As illustrated in Fig.~\ref{fig:framework}-\Romannum{2}, S1Pert processes all three text streams through a parameter-shared, frozen $L$-layer Jev backbone~\cite{mapika2026decider4b} ($L = 32$, $d = 2560$) without autoregressive decoding.
Specifically, we split the shared encoder $\Phi = \Phi_{\mathrm{up}} \circ \Phi_{\mathrm{low}}$ at layer $L_s = 24$ into a lower single-stream encoder $\Phi_{\mathrm{low}}$ (layers $1$ to $L_s$) and an upper cross-stream encoder $\Phi_{\mathrm{up}}$ (layers $L_s+1$ to $L$).
In a single forward pass of $\Phi$ on each unique text prompt, we extract hidden states at both the intermediate split layer ($\tau = \mathrm{mid}$, via $\Phi_{\mathrm{mid}} = \Phi_{\mathrm{low}}$) and the final layer ($\tau = \mathrm{top}$, via $\Phi_{\mathrm{top}} = \Phi$).
Variable-length token states are then compressed into fixed-size landmarks via segment-pooling operators $\Gamma_{16}$ and $\Gamma_4$, where $\Gamma_{16}$ partitions a token sequence into 12 contiguous segments, averages each segment, and appends the last four token states, while $\Gamma_4$ averages three contiguous segments and appends the last token state.
Following the stream order in Fig.~\ref{fig:framework}-\Romannum{2}, this yields the intermediate and final-layer semantic landmarks:
\begin{align}
\boldsymbol{C}_i^\tau &= \Gamma_{16}\!\left(\Phi_\tau(T_i^C)\right) \in \R^{16\times d}, \label{eq:condition-landmarks} \\
\boldsymbol{Q}_j^\tau &= \Gamma_{16}\!\left(\Phi_\tau(T_j^Q)\right) \in \R^{16\times d}, \label{eq:query-landmarks} \\
\boldsymbol{E}_{ij}^\tau &= \Gamma_4\!\left(\Phi_\tau(T_{ij}^E)\right) \in \R^{4\times d}. \label{eq:evidence-landmarks}
\end{align}

Although Eqs.~\eqref{eq:condition-landmarks}--\eqref{eq:evidence-landmarks} encode each stream independently to preserve role-specific semantics, capturing cross-stream interactions for each pair $(i,j)$ without concatenating full-length prompts requires a secondary fusion pass.
To achieve this, S1Pert concatenates the intermediate landmarks into a compact 36-token sequence $\boldsymbol{Z}_{ij}$ and passes it through the upper eight layers $\Phi_{\mathrm{up}}$:
\begin{align}
\boldsymbol{Z}_{ij} &= \big[\boldsymbol{C}_i^{\mathrm{mid}};\; \boldsymbol{Q}_j^{\mathrm{mid}};\; \boldsymbol{E}_{ij}^{\mathrm{mid}}\big] \in \R^{36\times d}, \label{eq:joint-sequence} \\
\widehat{\boldsymbol{Z}}_{ij} &= \big[\widehat{\boldsymbol{C}}_{ij};\; \widehat{\boldsymbol{Q}}_{ij};\; \widehat{\boldsymbol{E}}_{ij}\big] = \Phi_{\mathrm{up}}(\boldsymbol{Z}_{ij}),
\end{align}
where $\widehat{\boldsymbol{C}}_{ij} \in \R^{16\times d}$, $\widehat{\boldsymbol{Q}}_{ij} \in \R^{16\times d}$, and $\widehat{\boldsymbol{E}}_{ij} \in \R^{4\times d}$ denote the resulting contextualized stream blocks.
Let $\operatorname{Mean}(\boldsymbol{A}) = \frac{1}{m}\sum_{t=1}^m \boldsymbol{A}_{t,:} \in \R^{1\times d}$ denote the token-wise average of $\boldsymbol{A} \in \R^{m\times d}$.
Averaging each stream block alongside the full sequence $\widehat{\boldsymbol{Z}}_{ij}$ produces a four-token joint summary:
\begin{equation}
\begin{aligned}
\boldsymbol{J}_{ij} = \Big[
&\operatorname{Mean}(\widehat{\boldsymbol{C}}_{ij});\;
\operatorname{Mean}(\widehat{\boldsymbol{Q}}_{ij}); \\
&\operatorname{Mean}(\widehat{\boldsymbol{E}}_{ij});\;
\operatorname{Mean}(\widehat{\boldsymbol{Z}}_{ij})\Big] \in \R^{4\times d}.
\end{aligned}
\label{eq:joint-landmarks}
\end{equation}
To support evidence-dropout training, a prior-free counterpart of $\boldsymbol{J}_{ij}$ is also precomputed by zeroing $\boldsymbol{E}_{ij}^{\mathrm{mid}}$ and excluding its four positions from attention and averaging.

Because $T_{ij}^E$ takes values from only 109 distinct evidence prompts across the benchmark grid, Eqs.~\eqref{eq:condition-landmarks}--\eqref{eq:evidence-landmarks} evaluate full-length prompts on only $|\mathcal{T}_\ell| + |\mathcal{G}_\ell| + 109$ unique single-stream inputs, while Eq.~\eqref{eq:joint-sequence} reduces pair-level fusion to lightweight 36-token passes through $\Phi_{\mathrm{up}}$.
All final-layer single-stream landmarks ($\boldsymbol{C}_i^{\mathrm{top}}, \boldsymbol{Q}_j^{\mathrm{top}}, \boldsymbol{E}_{ij}^{\mathrm{top}}$) and joint summaries ($\boldsymbol{J}_{ij}$) are precomputed once and cached on disk, allowing online training and inference to directly load cached semantic tensors alongside the continuous covariates $(\boldsymbol{x}_i^C, \boldsymbol{x}_j^Q, \boldsymbol{x}_{ij}^E)$ without invoking the Jev backbone.

\subsection{Hierarchical Conditional Decoding}

During online execution, the conditional decoder ($D = 512$, $n_h = 8$ attention heads) maps the cached semantic landmarks and continuous covariates from Fig.~\ref{fig:framework}-\Romannum{2} into a shared $D$-dimensional space.
Along the semantic pathway, stream-specific linear projections with additive role embeddings map the cached final-layer landmarks $(\boldsymbol{C}_i^{\mathrm{top}}, \boldsymbol{Q}_j^{\mathrm{top}}, \boldsymbol{E}_{ij}^{\mathrm{top}})$ into $\widetilde{\boldsymbol{C}}_i \in \R^{16\times D}$, $\widetilde{\boldsymbol{Q}}_j \in \R^{16\times D}$, and $\widetilde{\boldsymbol{E}}_{ij} \in \R^{4\times D}$.
In parallel, three trainable stream-specific two-layer MLPs with GELU activations, denoted as $N_C(\cdot)$, $N_Q(\cdot)$, and $N_E(\cdot)$, project the continuous covariates $\boldsymbol{x}_i^C \in \R^4$, $\boldsymbol{x}_j^Q \in \R^4$, and $\boldsymbol{x}_{ij}^E \in \R^{12}$ into single-token representations $\tilde{\boldsymbol{x}}_i^C = N_C(\boldsymbol{x}_i^C) \in \R^{1\times D}$, $\tilde{\boldsymbol{x}}_j^Q = N_Q(\boldsymbol{x}_j^Q) \in \R^{1\times D}$, and $\tilde{\boldsymbol{x}}_{ij}^E = N_E(\boldsymbol{x}_{ij}^E) \in \R^{1\times D}$ (Fig.~\ref{fig:framework}-\Romannum{2}).

Conditioning downstream gene readouts on the upstream perturbation concatenates these projected tokens into a 17-token target memory $\bar{\boldsymbol{C}}_i$ and a 23-token query--evidence sequence $\bar{\boldsymbol{X}}_{ij}$ ending with a learnable readout token $\boldsymbol{r}_0 \in \R^{1\times D}$:
\begin{align}
    \bar{\boldsymbol{C}}_i &= \big[\tilde{\boldsymbol{x}}_i^C;\; \widetilde{\boldsymbol{C}}_i\big] \in \R^{17\times D}, \\
    \bar{\boldsymbol{X}}_{ij} &= \big[\tilde{\boldsymbol{x}}_j^Q;\; \widetilde{\boldsymbol{Q}}_j;\; \tilde{\boldsymbol{x}}_{ij}^E;\; \widetilde{\boldsymbol{E}}_{ij};\; \boldsymbol{r}_0\big] \in \R^{23\times D}.
\end{align}
Two stacked Pre-LN Transformer decoder blocks then update $\bar{\boldsymbol{X}}_{ij}$ (Fig.~\ref{fig:framework}-\Romannum{3}).
Each block applies masked self-attention over $\bar{\boldsymbol{X}}_{ij}$, cross-attention using $\bar{\boldsymbol{C}}_i$ as keys and values, and a GELU feed-forward network, yielding the normalized readout state $\boldsymbol{r}_{ij} \in \R^D$ at the final token position.
The cached upper-layer joint summary $\boldsymbol{J}_{ij}$ from Eq.~\eqref{eq:joint-landmarks} is similarly projected into $\bar{\boldsymbol{J}}_{ij} \in \R^{4\times D}$ and fused with $\boldsymbol{r}_{ij}$ via a multi-head cross-attention layer $\Attn_{\mathrm{joint}}(\cdot,\cdot)$ that takes $\boldsymbol{r}_{ij}$ as the query and $\bar{\boldsymbol{J}}_{ij}$ as keys and values:
\begin{equation}
    \boldsymbol{h}_{ij} = \boldsymbol{r}_{ij} + \Attn_{\mathrm{joint}}\!\left(\boldsymbol{r}_{ij},\, \bar{\boldsymbol{J}}_{ij}\right) \in \R^D,
    \label{eq:joint-attention}
\end{equation}
where the output projection of $\Attn_{\mathrm{joint}}$ is zero-initialized so that training starts from single-stream conditioning and gradually incorporates cross-stream interactions.

Because the dominant unchanged class ($y_{ij} = 1$) in perturbation screens easily overwhelms minority directional signals under a flat three-way classifier, the decoder decouples response detection from directional polarity.
Specifically, two linear heads map $\boldsymbol{h}_{ij}$ to a two-dimensional decision vector $[c_{ij}, d_{ij}]^{\mathsf{T}} \in \R^2$ comprising a change logit $c_{ij} \in \R$ and a conditional direction logit $d_{ij} \in \R$:
\begin{align}
c_{ij} &= \boldsymbol{w}_c^{\mathsf{T}}\boldsymbol{h}_{ij} + b_c, \label{eq:c-logit} \\
d_{ij} &= \boldsymbol{w}_d^{\mathsf{T}}\boldsymbol{h}_{ij} + b_d, \label{eq:d-logit}
\end{align}
where $\boldsymbol{w}_c, \boldsymbol{w}_d \in \R^D$ and $b_c, b_d \in \R$ are learnable parameters.
Applying the logistic sigmoid $\sig(\cdot)$ to the decision vector defines the stage-level change probability $P(\mathrm{change}) = \sig(c_{ij})$ and conditional upregulation probability $P(\mathrm{up}\mid\mathrm{change}) = \sig(d_{ij})$.
Following the hierarchical decision tree in Fig.~\ref{fig:framework}-\Romannum{3}, S1Pert first evaluates whether expression changes (\textbf{Unchanged} vs.\ \textbf{Changed}) and then resolves directional polarity (\textbf{Upregulated} vs.\ \textbf{Downregulated}), yielding the factorized three-class posterior probabilities:
\begin{align}
    \hat{p}_{ij,1} &= P(\mathrm{same}) = 1 - \sig(c_{ij}), \label{eq:prob-same} \\
    \hat{p}_{ij,0} &= P(\mathrm{up}) = \sig(c_{ij})\sig(d_{ij}), \label{eq:prob-up} \\
    \hat{p}_{ij,2} &= P(\mathrm{down}) = \sig(c_{ij})\big[1 - \sig(d_{ij})\big]. \label{eq:prob-down}
\end{align}
Online inference then predicts $\hat{y}_{ij} = \arg\max_{k\in\{0,1,2\}} \hat{p}_{ij,k}$ in a single forward pass.

\subsection{Training Objective}

To jointly learn response detection and target-specific directional polarity while suppressing historical gene-frequency shortcuts, the conditional decoder is optimized under the three training mechanisms shown at the bottom of Fig.~\ref{fig:framework}-\Romannum{3}: synchronized (random) evidence dropout, binary cross-entropy loss, and cross-target pairwise ranking loss.

\noindent\textbf{Synchronized Evidence Dropout.}
During training, each instance draws a retention gate $\xi_{ij} \sim \mathrm{Bernoulli}(1 - p_{\mathrm{ev}})$ with dropout rate $p_{\mathrm{ev}} = 0.10$, retaining all historical-evidence inputs when $\xi_{ij} = 1$ and masking $\widetilde{\boldsymbol{E}}_{ij}$, resetting $\boldsymbol{x}_{ij}^E$ to an uninformative prior $[1/3, 1/3, 1/3, 0, \ldots, 0]^{\mathsf{T}}$, and switching $\boldsymbol{J}_{ij}$ to its cached prior-free version when $\xi_{ij} = 0$.
Randomly dropping all historical evidence forces the decoder to infer transcriptional responses from target and query features alone when historical priors are weak or unavailable.

\noindent\textbf{Binary Cross-Entropy Loss.}
Let $\ind[\cdot]$ denote the indicator function, $z_{ij} = \ind[y_{ij} \neq 1]$ indicate expression change, $b_{ij} = \ind[y_{ij} = 0]$ indicate upregulation, and $\BCE(\cdot)$ denote the binary cross-entropy loss.
Over a training batch $\mathcal{B}$, the hierarchical classification loss is computed on the decision logits $(c_{ij}, d_{ij})$ as
\begin{equation}
    \mathcal{L}_{\mathrm{main}} = \frac{1}{|\mathcal{B}|}\sum_{(i,j)\in\mathcal{B}}\Big[\BCE(c_{ij}, z_{ij}) + z_{ij}\BCE(d_{ij}, b_{ij})\Big],
    \label{eq:main-loss}
\end{equation}
where multiplying the second term by $z_{ij}$ trains the polarity head $d_{ij}$ only on changed instances, preventing unchanged pairs from diluting directional supervision.

\noindent\textbf{Pairwise Ranking Loss.}
Fitting $\mathcal{L}_{\mathrm{main}}$ alone allows the model to repeat each gene's historical propensity $\boldsymbol{p}_{ij}^{\mathrm{ev}}$ while ignoring the perturbed target $i$.
For the same query gene $j$ across two distinct training targets $(i, i')$, S1Pert samples change versus no-change tuples $(i, i', j) \in \mathcal{P}^{\mathrm{cs}}$ ($y_{ij} \neq 1,\; y_{i'j} = 1$) and upregulation versus downregulation tuples $(i, i', j) \in \mathcal{P}^{\mathrm{ud}}$ ($y_{ij} = 0,\; y_{i'j} = 2$), computing the margin ranking loss ($m_p = 0.05$, $[x]_+ = \max(0,x)$) as
\begin{equation}
\begin{aligned}
    \mathcal{L}_{\mathrm{rank}} &= \frac{1}{|\mathcal{P}^{\mathrm{cs}}|}\sum_{(i,i',j)\in\mathcal{P}^{\mathrm{cs}}}\big[m_p - c_{ij} + c_{i'j}\big]_+ \\
    &\quad + \frac{1}{|\mathcal{P}^{\mathrm{ud}}|}\sum_{(i,i',j)\in\mathcal{P}^{\mathrm{ud}}}\big[m_p - d_{ij} + d_{i'j}\big]_+,
\end{aligned}
\label{eq:pair-loss}
\end{equation}
where comparing targets $i$ and $i'$ on the same query gene $j$ forces the decoder to rank target-specific effects ($c_{ij} > c_{i'j}$ and $d_{ij} > d_{i'j}$) rather than relying on shared gene-level priors.
The total training objective is $\mathcal{L} = \mathcal{L}_{\mathrm{main}} + \lambda_p \mathcal{L}_{\mathrm{rank}}$ with $\lambda_p = 0.02$.

\section{Experiments}
\subsection{Experimental Settings}
\label{sec:experimental_settings}


\begin{table*}[!t]
\centering
\caption{Performance on four genetic perturbation datasets.}
\label{tab:s1pert_main_comparison}
\begingroup
\fontsize{8}{10}\selectfont
\setlength{\tabcolsep}{1.0pt}
\renewcommand{\arraystretch}{1.00}
\definecolor{RankFirst}{HTML}{F6CCCC}
\definecolor{RankSecond}{HTML}{FCE4CB}
\newcommand{\TStd}[1]{{\fontsize{6}{7}\selectfont\ensuremath{\mathord{\pm}}#1}}
\newcommand{\TValue}[2]{#1\TStd{#2}}
\newcommand{\TBest}[2]{\cellcolor{RankFirst}\textbf{#1}\TStd{#2}}
\newcommand{\TSecond}[2]{\cellcolor{RankSecond}\underline{#1}\TStd{#2}}
\begin{tabular*}{\textwidth}{@{\extracolsep{\fill}}l*{11}{c}@{}}
\toprule
\multirow{2}{*}{\textbf{Method}}
& \multicolumn{2}{c}{\textbf{Overall}}
& \multicolumn{3}{c}{\textbf{Upregulated}}
& \multicolumn{3}{c}{\textbf{Downregulated}}
& \multicolumn{3}{c}{\textbf{Unchanged}} \\
\cmidrule(lr){2-3}\cmidrule(lr){4-6}\cmidrule(lr){7-9}\cmidrule(lr){10-12}
& \shortstack{Macro\\F1-score$\uparrow$}
& \shortstack{Balanced\\Accuracy$\uparrow$}
& Precision$\uparrow$ & Recall$\uparrow$ & F1-score$\uparrow$
& Precision$\uparrow$ & Recall$\uparrow$ & F1-score$\uparrow$
& Precision$\uparrow$ & Recall$\uparrow$ & F1-score$\uparrow$ \\
\midrule
\multicolumn{12}{l}{\textbf{Dataset: HepG2}} \\
GenePT-Gene
& \TValue{0.694}{0.015} & \TValue{0.690}{0.015}
& \TValue{0.687}{0.035} & \TValue{0.522}{0.020} & \TValue{0.593}{0.017}
& \TValue{0.676}{0.041} & \TValue{0.741}{0.033} & \TValue{0.705}{0.012}
& \TValue{0.767}{0.059} & \TValue{0.807}{0.050} & \TValue{0.784}{0.024} \\

GenePT-Prot
& \TValue{0.751}{0.017} & \TValue{0.747}{0.015}
& \TValue{0.744}{0.024} & \TValue{0.620}{0.023} & \TValue{0.676}{0.013}
& \TValue{0.724}{0.043} & \TValue{0.789}{0.013} & \TValue{0.755}{0.023}
& \TSecond{0.818}{0.028} & \TValue{0.831}{0.051}
& \TSecond{0.823}{0.023} \\

Qwen3-4B-LoRA
& \TValue{0.618}{0.021} & \TValue{0.617}{0.027}
& \TValue{0.517}{0.032} & \TValue{0.456}{0.100} & \TValue{0.479}{0.044}
& \TValue{0.575}{0.045} & \TValue{0.642}{0.084} & \TValue{0.602}{0.028}
& \TValue{0.809}{0.070} & \TValue{0.753}{0.094} & \TValue{0.773}{0.036} \\

GEARS
& \TValue{0.667}{0.011} & \TValue{0.667}{0.014}
& \TValue{0.796}{0.028} & \TSecond{0.771}{0.054}
& \TSecond{0.782}{0.033}
& \TValue{0.667}{0.037} & \TValue{0.548}{0.073} & \TValue{0.598}{0.036}
& \TValue{0.576}{0.037} & \TValue{0.682}{0.089} & \TValue{0.622}{0.042} \\

scGPT
& \TValue{0.515}{0.030} & \TValue{0.517}{0.035}
& \TValue{0.481}{0.034} & \TValue{0.463}{0.117} & \TValue{0.467}{0.071}
& \TValue{0.638}{0.085} & \TValue{0.498}{0.066} & \TValue{0.553}{0.032}
& \TValue{0.478}{0.046} & \TValue{0.589}{0.045} & \TValue{0.525}{0.021} \\

STATE
& \TValue{0.709}{0.034} & \TValue{0.704}{0.036}
& \TValue{0.757}{0.025} & \TValue{0.650}{0.062} & \TValue{0.698}{0.043}
& \TValue{0.746}{0.054} & \TValue{0.776}{0.025} & \TValue{0.760}{0.031}
& \TValue{0.655}{0.050} & \TValue{0.687}{0.043} & \TValue{0.670}{0.039} \\

ALIGNED
& \TSecond{0.791}{0.029} & \TSecond{0.789}{0.025}
& \TSecond{0.806}{0.063} & \TValue{0.738}{0.018} & \TValue{0.769}{0.032}
& \TValue{0.816}{0.051} & \TSecond{0.803}{0.018}
& \TSecond{0.809}{0.025}
& \TValue{0.771}{0.038} & \TValue{0.827}{0.067} & \TValue{0.796}{0.030} \\

SUMMER
& \TValue{0.724}{0.018} & \TValue{0.714}{0.014}
& \TValue{0.800}{0.031} & \TValue{0.645}{0.040} & \TValue{0.713}{0.024}
& \TBest{0.849}{0.024} & \TValue{0.624}{0.019} & \TValue{0.719}{0.007}
& \TValue{0.643}{0.054} & \TBest{0.875}{0.023} & \TValue{0.739}{0.031} \\

\textbf{S1Pert (Ours)}
& \TBest{0.850}{0.045} & \TBest{0.851}{0.042}
& \TBest{0.860}{0.045} & \TBest{0.817}{0.045}
& \TBest{0.837}{0.033}
& \TSecond{0.829}{0.099} & \TBest{0.874}{0.024}
& \TBest{0.847}{0.044}
& \TBest{0.883}{0.023} & \TSecond{0.861}{0.120}
& \TBest{0.867}{0.060} \\

\midrule

\multicolumn{12}{l}{\textbf{Dataset: Jurkat}} \\
GenePT-Gene
& \TValue{0.543}{0.029} & \TValue{0.531}{0.044}
& \TValue{0.562}{0.139} & \TValue{0.321}{0.080} & \TValue{0.393}{0.060}
& \TValue{0.404}{0.116} & \TValue{0.376}{0.125} & \TValue{0.360}{0.054}
& \TValue{0.861}{0.043} & \TValue{0.896}{0.072} & \TValue{0.875}{0.015} \\

GenePT-Prot
& \TValue{0.578}{0.057} & \TValue{0.567}{0.054}
& \TValue{0.584}{0.146} & \TValue{0.380}{0.107} & \TValue{0.447}{0.109}
& \TValue{0.429}{0.135} & \TSecond{0.437}{0.106} & \TValue{0.409}{0.071}
& \TSecond{0.874}{0.026} & \TValue{0.886}{0.083} & \TValue{0.878}{0.031} \\

Qwen3-4B-LoRA
& \TValue{0.430}{0.027} & \TValue{0.439}{0.024}
& \TValue{0.367}{0.249} & \TValue{0.180}{0.193} & \TValue{0.187}{0.121}
& \TValue{0.334}{0.162} & \TValue{0.211}{0.186} & \TValue{0.224}{0.146}
& \TValue{0.838}{0.019} & \TSecond{0.927}{0.066} & \TValue{0.879}{0.031} \\

GEARS
& \TValue{0.560}{0.059} & \TValue{0.542}{0.054}
& \TSecond{0.648}{0.172} & \TValue{0.500}{0.079} & \TValue{0.554}{0.081}
& \TValue{0.299}{0.090} & \TValue{0.245}{0.091} & \TValue{0.269}{0.092}
& \TValue{0.833}{0.031} & \TValue{0.881}{0.014} & \TValue{0.856}{0.022} \\

scGPT
& \TValue{0.393}{0.015} & \TValue{0.400}{0.016}
& \TValue{0.172}{0.048} & \TValue{0.227}{0.062} & \TValue{0.189}{0.036}
& \TValue{0.241}{0.046} & \TValue{0.176}{0.050} & \TValue{0.197}{0.030}
& \TValue{0.792}{0.027} & \TValue{0.798}{0.051} & \TValue{0.794}{0.020} \\

STATE
& \TValue{0.561}{0.027} & \TValue{0.546}{0.026}
& \TValue{0.529}{0.092} & \TValue{0.405}{0.051} & \TValue{0.457}{0.060}
& \TValue{0.388}{0.050} & \TValue{0.366}{0.049} & \TValue{0.375}{0.036}
& \TValue{0.838}{0.027} & \TValue{0.868}{0.010} & \TValue{0.852}{0.009} \\

ALIGNED
& \TSecond{0.638}{0.042} & \TSecond{0.618}{0.011}
& \TValue{0.637}{0.142} & \TSecond{0.509}{0.014}
& \TSecond{0.559}{0.064}
& \TValue{0.523}{0.132} & \TValue{0.432}{0.027}
& \TSecond{0.464}{0.048}
& \TValue{0.873}{0.020} & \TValue{0.914}{0.053}
& \TSecond{0.892}{0.019} \\

SUMMER
& \TValue{0.583}{0.012} & \TValue{0.545}{0.008}
& \TValue{0.635}{0.082} & \TValue{0.432}{0.022} & \TValue{0.513}{0.041}
& \TBest{0.532}{0.073} & \TValue{0.260}{0.028} & \TValue{0.346}{0.013}
& \TValue{0.842}{0.025} & \TBest{0.942}{0.017} & \TValue{0.889}{0.011} \\

\textbf{S1Pert (Ours)}
& \TBest{0.698}{0.071} & \TBest{0.698}{0.039}
& \TBest{0.707}{0.094} & \TBest{0.614}{0.060}
& \TBest{0.657}{0.073}
& \TSecond{0.529}{0.187} & \TBest{0.590}{0.053}
& \TBest{0.537}{0.097}
& \TBest{0.913}{0.014} & \TValue{0.890}{0.085}
& \TBest{0.900}{0.042} \\

\midrule

\multicolumn{12}{l}{\textbf{Dataset: RPE1}} \\
GenePT-Gene
& \TValue{0.645}{0.015} & \TValue{0.638}{0.013}
& \TValue{0.573}{0.022} & \TValue{0.438}{0.025} & \TValue{0.495}{0.014}
& \TValue{0.674}{0.020} & \TValue{0.770}{0.014} & \TValue{0.718}{0.013}
& \TValue{0.739}{0.028} & \TValue{0.707}{0.049} & \TValue{0.722}{0.035} \\

GenePT-Prot
& \TValue{0.704}{0.030} & \TValue{0.697}{0.023}
& \TValue{0.656}{0.056} & \TValue{0.525}{0.015} & \TValue{0.582}{0.021}
& \TValue{0.710}{0.046} & \TValue{0.821}{0.006} & \TValue{0.760}{0.026}
& \TSecond{0.802}{0.009} & \TValue{0.745}{0.079}
& \TSecond{0.770}{0.045} \\

Qwen3-4B-LoRA
& \TValue{0.531}{0.022} & \TValue{0.533}{0.022}
& \TValue{0.400}{0.143} & \TValue{0.212}{0.019} & \TValue{0.268}{0.021}
& \TValue{0.597}{0.058} & \TValue{0.707}{0.129} & \TValue{0.639}{0.042}
& \TValue{0.720}{0.093} & \TValue{0.680}{0.117} & \TValue{0.688}{0.022} \\

GEARS
& \TValue{0.662}{0.035} & \TValue{0.659}{0.043}
& \TValue{0.677}{0.022} & \TValue{0.627}{0.103} & \TValue{0.647}{0.060}
& \TValue{0.694}{0.012} & \TValue{0.716}{0.049} & \TValue{0.704}{0.029}
& \TValue{0.639}{0.054} & \TValue{0.634}{0.040} & \TValue{0.635}{0.034} \\

scGPT
& \TValue{0.506}{0.023} & \TValue{0.500}{0.021}
& \TValue{0.429}{0.037} & \TValue{0.287}{0.060} & \TValue{0.342}{0.050}
& \TValue{0.641}{0.054} & \TValue{0.598}{0.069} & \TValue{0.614}{0.026}
& \TValue{0.520}{0.031} & \TValue{0.614}{0.058} & \TValue{0.561}{0.024} \\

STATE
& \TValue{0.680}{0.017} & \TValue{0.673}{0.020}
& \TValue{0.675}{0.030} & \TValue{0.585}{0.069} & \TValue{0.625}{0.043}
& \TValue{0.738}{0.026} & \TValue{0.752}{0.059} & \TValue{0.743}{0.020}
& \TValue{0.663}{0.042} & \TValue{0.683}{0.026} & \TValue{0.672}{0.019} \\

ALIGNED
& \TSecond{0.727}{0.016} & \TSecond{0.728}{0.010}
& \TSecond{0.688}{0.037} & \TSecond{0.660}{0.010}
& \TSecond{0.674}{0.018}
& \TValue{0.732}{0.030} & \TSecond{0.854}{0.014}
& \TSecond{0.788}{0.016}
& \TValue{0.777}{0.017} & \TValue{0.670}{0.038} & \TValue{0.719}{0.018} \\

SUMMER
& \TValue{0.654}{0.024} & \TValue{0.638}{0.023}
& \TValue{0.660}{0.024} & \TValue{0.498}{0.044} & \TValue{0.566}{0.028}
& \TSecond{0.777}{0.024} & \TValue{0.633}{0.039} & \TValue{0.697}{0.023}
& \TValue{0.632}{0.043} & \TBest{0.782}{0.027} & \TValue{0.698}{0.022} \\

\textbf{S1Pert (Ours)}
& \TBest{0.785}{0.024} & \TBest{0.791}{0.017}
& \TBest{0.726}{0.063} & \TBest{0.761}{0.036}
& \TBest{0.740}{0.023}
& \TBest{0.785}{0.033} & \TBest{0.860}{0.025}
& \TBest{0.820}{0.024}
& \TBest{0.842}{0.027} & \TSecond{0.752}{0.057}
& \TBest{0.793}{0.029} \\

\midrule

\multicolumn{12}{l}{\textbf{Dataset: K562}} \\
GenePT-Gene
& \TValue{0.552}{0.024} & \TValue{0.533}{0.031}
& \TValue{0.551}{0.054} & \TValue{0.409}{0.069} & \TValue{0.466}{0.051}
& \TValue{0.404}{0.072} & \TValue{0.279}{0.055} & \TValue{0.322}{0.027}
& \TValue{0.832}{0.029} & \TValue{0.910}{0.032} & \TValue{0.868}{0.009} \\

GenePT-Prot
& \TValue{0.619}{0.007} & \TValue{0.596}{0.014}
& \TValue{0.631}{0.028} & \TValue{0.506}{0.050} & \TValue{0.561}{0.036}
& \TValue{0.476}{0.067} & \TValue{0.358}{0.026} & \TValue{0.405}{0.010}
& \TValue{0.861}{0.019} & \TValue{0.924}{0.025} & \TValue{0.891}{0.016} \\

Qwen3-4B-LoRA
& \TValue{0.487}{0.042} & \TValue{0.489}{0.024}
& \TValue{0.489}{0.118} & \TValue{0.407}{0.154} & \TValue{0.412}{0.034}
& \TValue{0.258}{0.170} & \TValue{0.174}{0.110} & \TValue{0.197}{0.116}
& \TValue{0.826}{0.033} & \TValue{0.885}{0.061} & \TValue{0.853}{0.030} \\

GEARS
& \TValue{0.563}{0.023} & \TValue{0.574}{0.016}
& \TValue{0.631}{0.081} & \TSecond{0.713}{0.065} & \TValue{0.663}{0.016}
& \TValue{0.233}{0.062} & \TValue{0.155}{0.041} & \TValue{0.183}{0.040}
& \TValue{0.833}{0.030} & \TValue{0.854}{0.045} & \TValue{0.843}{0.030} \\

scGPT
& \TValue{0.368}{0.025} & \TValue{0.368}{0.024}
& \TValue{0.194}{0.069} & \TValue{0.213}{0.051} & \TValue{0.202}{0.060}
& \TValue{0.177}{0.042} & \TValue{0.128}{0.032} & \TValue{0.148}{0.036}
& \TValue{0.746}{0.030} & \TValue{0.763}{0.028} & \TValue{0.754}{0.028} \\

STATE
& \TValue{0.554}{0.019} & \TValue{0.548}{0.016}
& \TValue{0.544}{0.065} & \TValue{0.507}{0.093} & \TValue{0.522}{0.070}
& \TValue{0.337}{0.055} & \TValue{0.297}{0.052} & \TValue{0.313}{0.042}
& \TValue{0.817}{0.024} & \TValue{0.840}{0.033} & \TValue{0.828}{0.024} \\

ALIGNED
& \TSecond{0.705}{0.009} & \TSecond{0.689}{0.007}
& \TSecond{0.718}{0.052} & \TValue{0.688}{0.016}
& \TSecond{0.701}{0.023}
& \TSecond{0.580}{0.074} & \TSecond{0.445}{0.019}
& \TSecond{0.501}{0.014}
& \TSecond{0.893}{0.020} & \TSecond{0.935}{0.015}
& \TSecond{0.913}{0.006} \\

SUMMER
& \TValue{0.627}{0.022} & \TValue{0.591}{0.022}
& \TValue{0.701}{0.049} & \TValue{0.550}{0.033} & \TValue{0.616}{0.033}
& \TValue{0.554}{0.066} & \TValue{0.285}{0.041} & \TValue{0.375}{0.045}
& \TValue{0.845}{0.024} & \TBest{0.938}{0.009} & \TValue{0.889}{0.012} \\

\textbf{S1Pert (Ours)}
& \TBest{0.769}{0.009} & \TBest{0.769}{0.007}
& \TBest{0.770}{0.030} & \TBest{0.775}{0.006}
& \TBest{0.772}{0.018}
& \TBest{0.619}{0.084} & \TBest{0.596}{0.042}
& \TBest{0.602}{0.020}
& \TBest{0.931}{0.016} & \TValue{0.935}{0.021}
& \TBest{0.933}{0.009} \\

\bottomrule
\end{tabular*}
\par\vspace{3pt}
\raggedright\fontsize{7}{8}\selectfont
Results are reported as mean $\pm$ standard deviation over five
target-held-out splits.
Within each dataset and metric, light red shading with bold means
indicates the best result; light orange shading with underlined
means indicates the second-best result.
Ties are broken by smaller standard deviation.
$\uparrow$: higher is better.\par
\endgroup
\end{table*}

\textbf{Benchmark Datasets.}
Experiments are conducted on four human genetic perturbation datasets covering HepG2, Jurkat, RPE1, and K562~\cite{nadig2025transcriptome,replogle2022mapping}.
Applying the statistical label adjudication protocol in \textbf{Algorithm~1} ($\rho = 1.2$, $\alpha = 0.05$) and excluding statistically inconclusive pairs ($y_{ij} = \bot$) yields the resolved three-class benchmark datasets summarized in Table~\ref{tab:dataset_statistics}.
Each cell line is modeled independently under five repeated target-held-out splits, partitioning perturbation targets $\mathcal{T}_\ell$ into mutually exclusive training, validation, and test sets at a $7{:}1{:}2$ ratio so that all query-gene instances of a given target reside in a single partition.
To ensure a strictly fair comparison, S1Pert and all compared baselines are trained and evaluated on the exact same adjudicated labels, target splits, and resolved target--gene pairs.

\begin{table}[t]
    \centering
    \caption{Label distribution across four genetic perturbation datasets.}
    \label{tab:dataset_statistics}
    \small
    \begin{tabular}{lrrr}
        \toprule
        Dataset & Upregulated & Downregulated & Unchanged \\
        \midrule
        HepG2          & 95,183  & 154,197 & 168,307 \\
        Jurkat         & 30,877  & 52,592  & 323,094 \\
        RPE1           & 117,388 & 303,859 & 316,561 \\
        K562 & 101,304 & 84,161  & 533,703 \\
        \bottomrule
    \end{tabular}
\end{table}

\noindent\textbf{Compared Methods.}
To validate the effectiveness of our proposed S1Pert model, we compare its performance with representative gene representation models, perturbation response predictors, and general-purpose language model baselines. These comparative methods can be broadly categorized into three groups:

\noindent\textit{(1) Pre-trained Language and Gene Embedding Models:}
\begin{itemize}
    \item \textbf{GenePT-Gene}~\cite{chen2025genept,chen2024geneptembeddings}, which is a semantic embedding model encoding textual gene descriptions into dense representations for response classification.
    \item \textbf{GenePT-Prot}~\cite{chen2025genept,chen2024geneptembeddings}, which is a joint gene--protein embedding variant capturing functional semantics across biological text and protein modalities.
    \item \textbf{Qwen3-4B-LoRA~\cite{qwen3technicalreport}}, which is a parameter-efficient fine-tuning baseline adapting the Qwen3-4B Transformer with low-rank adapters and a discriminative classification head.
\end{itemize}

\noindent\textit{(2) Continuous Perturbation Profile Predictors:}
\begin{itemize}
    \item \textbf{GEARS}~\cite{roohani2024gears}, which is a graph neural network integrating gene co-expression and Gene Ontology knowledge graphs to regress post-perturbation expression profiles.
    \item \textbf{scGPT}~\cite{cui2024scgpt}, which is a pre-trained single-cell foundation Transformer adapted for perturbation-conditioned transcriptomic shift prediction.
\item \textbf{STATE}~\cite{State2026}, which combines State Embedding (SE) for learning cell-state representations with State Transition (ST) for modeling perturbation effects across cell populations to improve response prediction across diverse cellular contexts.
\end{itemize}

\noindent\textit{(3) Neuro-Symbolic and Knowledge-Grounded Reasoning Models:}
\begin{itemize}
    \item \textbf{ALIGNED}~\cite{xiang2026aligned}, which is a neuro-symbolic abductive learning framework aligning neural perturbation predictions with symbolic regulatory knowledge.
    \item \textbf{SUMMER}~\cite{wu2025perturbqa}, which is a knowledge-grounded prompting framework synthesizing biological text and training priors for directional response prediction.

\end{itemize}
All compared methods undergo task-specific adaptation for three-class response prediction. For continuous perturbation profile predictors, predicted expression changes are mapped to response classes using thresholds calibrated on the validation set.

\noindent\textbf{Evaluation Metrics.}
Classification performance is evaluated using Macro-F1, balanced accuracy (BA), and class-wise Precision, Recall, and F1-score. All methods are evaluated on the same five random target-held-out splits for each cell line, with results reported as mean $\pm$ sample standard deviation. The best and second-best results are highlighted in bold and underlined, respectively.

\noindent\textbf{Implementation Details.}
S1Pert is implemented in PyTorch and trained on a single NVIDIA RTX PRO 6000 GPU. We instantiate the Jev-style System-One encoder using the open-source DECIDER-4B model~\cite{mapika2026decider4b}. The backbone is frozen, and its representations are precomputed and cached. The task-specific decoder is trained from scratch using AdamW with a learning rate of $10^{-4}$, weight decay of $10^{-2}$, and a batch size of 4,096 perturbation--query gene pairs. Training runs for up to 100 epochs, with a minimum of 10 epochs and early stopping after 10 epochs without a validation Macro-F1 improvement exceeding $10^{-4}$. The checkpoint with the highest validation Macro-F1 is selected for testing. Hyperparameters are fixed across cell lines and the five random target-held-out splits.

\begin{table}[!htbp]
\centering
\caption{Ablation study and target-permutation diagnostic on K562.}
\label{tab:ablation_summary}
\begingroup
\fontsize{8}{10}\selectfont
\setlength{\tabcolsep}{1pt}
\renewcommand{\arraystretch}{1.10}
\definecolor{RankFirst}{HTML}{F6CCCC}
\definecolor{RankSecond}{HTML}{FCE4CB}
\newcommand{\TValue}[2]{\shortstack[c]{#1\\{\fontsize{7}{8}\selectfont$\pm$#2}}}
\newcommand{\TBest}[2]{\cellcolor{RankFirst}\TValue{\textbf{#1}}{#2}}
\newcommand{\TSecond}[2]{\cellcolor{RankSecond}\TValue{\underline{#1}}{#2}}
\begin{tabular*}{\columnwidth}{@{\extracolsep{\fill}}l*{4}{c}@{}}
\toprule
\textbf{Variant}
& \shortstack{Macro\\F1-score$\uparrow$}
& \shortstack{Upregulated\\F1-score$\uparrow$}
& \shortstack{Unchanged\\F1-score$\uparrow$}
& \shortstack{Downregulated\\F1-score$\uparrow$} \\
\midrule
\textbf{S1Pert (full)}
& \TBest{0.769}{0.009} & \TBest{0.772}{0.018}
& \TBest{0.933}{0.009} & \TBest{0.602}{0.020} \\
\midrule
\texttt{NO\_Jev}
& \TValue{0.739}{0.016} & \TValue{0.755}{0.028}
& \TValue{0.920}{0.009} & \TValue{0.542}{0.022} \\
\addlinespace[2pt]
\texttt{NO\_JOINT}
& \TValue{0.755}{0.012} & \TValue{0.761}{0.025}
& \TValue{0.923}{0.008} & \TValue{0.582}{0.031} \\
\addlinespace[2pt]
\texttt{FLAT}
& \TSecond{0.764}{0.012} & \TSecond{0.770}{0.023}
& \TSecond{0.932}{0.006} & \TSecond{0.589}{0.033} \\
\addlinespace[2pt]
\texttt{NO\_RANK}
& \TValue{0.760}{0.012} & \TValue{0.769}{0.014}
& \TValue{0.926}{0.011} & \TValue{0.587}{0.030} \\
\addlinespace[2pt]
\texttt{NO\_MASK}
& \TValue{0.760}{0.007} & \TValue{0.767}{0.015}
& \TValue{0.927}{0.008} & \TValue{0.584}{0.023} \\
\midrule
\shortstack[l]{Target\\permutation}
& \TValue{0.703}{0.017} & \TValue{0.712}{0.023}
& \TValue{0.907}{0.014} & \TValue{0.489}{0.026} \\
\bottomrule
\end{tabular*}
\par\vspace{3pt}
\raggedright\fontsize{7}{8}\selectfont
Each cell shows the mean above $\pm$ sample standard deviation over five target-held-out splits. Light red/bold and light orange/underlined means indicate the best and second-best results within each metric, respectively. $\uparrow$: higher is better. Target permutation is evaluated at inference time; the other variants are retrained.\par
\endgroup
\end{table}

\subsection{Overall Performance}

As shown in Table~\ref{tab:s1pert_main_comparison}, S1Pert achieves the highest performance across all four benchmarks, reaching Macro-F1 scores of 0.850, 0.698, 0.785, and 0.769 and balanced accuracies (BA) of 0.851, 0.698, 0.791, and 0.769 on HepG2, Jurkat, RPE1, and K562, respectively.
Surpassing the strongest baseline in Table~\ref{tab:s1pert_main_comparison} by 5.8--6.4 percentage points in Macro-F1 and ranking first in class-wise F1 across all response states establishes the effectiveness of S1Pert for unseen-target response prediction.
In addition, several notable observations can be drawn from Table~\ref{tab:s1pert_main_comparison}.
ALIGNED outperforms the continuous profile predictors (GEARS, scGPT, and STATE) in Macro-F1 and secures the runner-up position in Table~\ref{tab:s1pert_main_comparison} across all cell lines, with scores of 0.791, 0.638, 0.727, and 0.705. SUMMER also outperforms these predictors on HepG2, Jurkat, and K562, but not on RPE1.
These results support the effectiveness of knowledge-grounded discrete classification under the evaluated protocol, although the relative performance varies by dataset and method.
The frozen representation models GenePT-Gene and GenePT-Prot consistently outperform Qwen3-4B-LoRA, with Macro-F1 gains of 6.5--17.3 percentage points. These results support the effectiveness of frozen functional representations with task-specific downstream decoding in this benchmark.

Unseen-target perturbation prediction is inherently challenged by severe class imbalance, where unchanged pairs dominate over minority directional shifts (Table~\ref{tab:dataset_statistics}).
The continuous profile predictors GEARS, scGPT, and STATE achieve Downregulated F1 scores of at most $0.313$ on K562 and $0.375$ on Jurkat, illustrating the difficulty of predicting this minority response class.
Although ALIGNED and SUMMER mitigate class imbalance through abductive alignment or generative prompting, their highest Downregulated F1 scores reach $0.464$ on Jurkat and $0.501$ on K562.
Mixing target, query, and prior information within a flat prompt or static graph obscures role-specific boundaries and lacks explicit cross-attention conditioning to isolate target-specific effects from gene-level tendencies.
In contrast, S1Pert overcomes these limitations by unifying role-decoupled multilevel Jev landmark caching with a hierarchical cross-attention decoder and deconfounded optimization, dynamically extracting target-specific regulatory shifts to achieve state-of-the-art performance.

\subsection{Ablation Study}
\label{sec:ablation_study}

\begin{figure*}[!t]
    \centering
    \includegraphics[width=\textwidth]{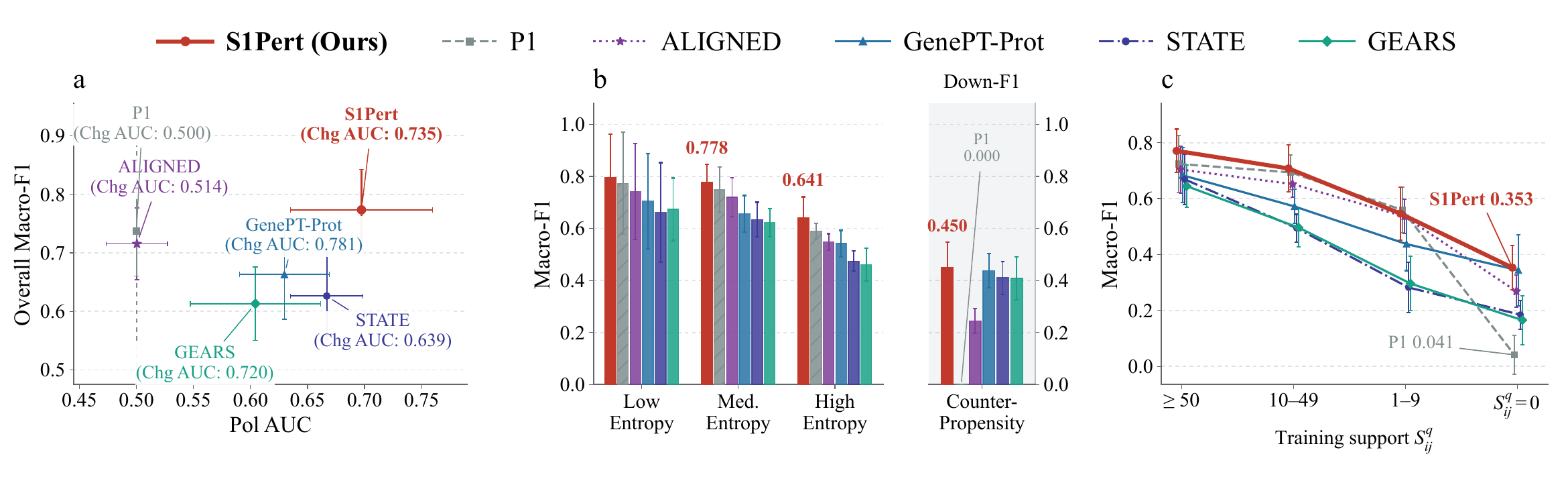}     
    \caption{Cross-target ranking and prior-stress evaluation across four datasets with five target-held-out splits each. (a) Overall pooled Macro-F1 versus within-gene Polarity AUC (\texttt{Pol AUC}) and Change AUC (\texttt{Chg AUC}, in parentheses). (b) Macro-F1 across out-of-fold prior-entropy tertiles (left) and Down-F1 under counter-propensity reversal ($y_{ij} \neq \arg\max_k p_{ij,k}^{\mathrm{ev}}$, right). (c) Macro-F1 across training-support regimes ($S_{ij}^q$), including zero-support cold start ($S_{ij}^q = 0$). Error bars denote $\pm 1$ standard deviation across target-held-out splits.}    
    \label{fig:prior_stress}
\end{figure*}

\begin{figure*}[!t]
    \centering
    \includegraphics[width=\textwidth]{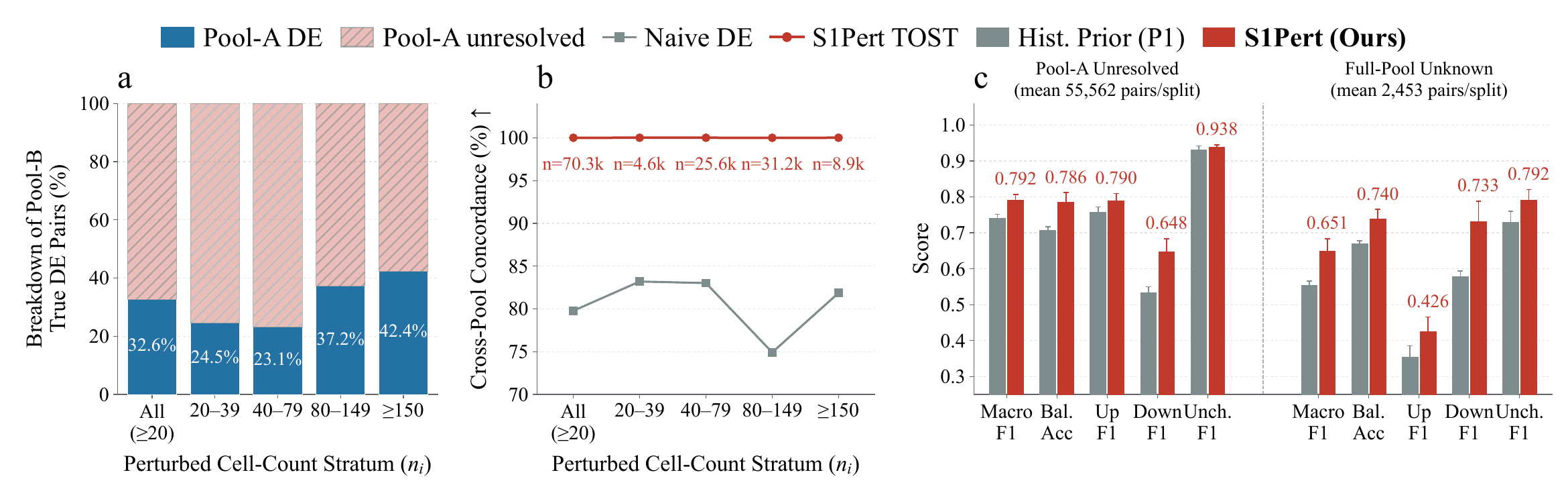}
    \caption{Disjoint cell-pool validation and boundary generalization on K562. (a) Adjudication of $76{,}047$ Pool-B reference DE pairs in lower-depth Pool A by cell-count stratum ($n_i$). Naive DE labels unresolved pairs as unchanged; S1Pert TOST marks them as inconclusive. (b) Cross-pool label concordance; $n$ denotes pairs resolved in both pools by S1Pert TOST. (c) S1Pert versus historical-prior baseline P1 on \texttt{Pool-A Unresolved} and \texttt{Full-Pool Unknown}, evaluated without retraining against Pool-B reference labels. Bars show means and error bars indicate $\pm 1$ standard deviation across five target-held-out splits.}
    
    \label{fig:label_boundary}
\end{figure*}

Five ablated model variants and an inference-time diagnostic protocol are evaluated on K562 to isolate the contribution of each design component in S1Pert.
Specifically, (1)~\texttt{NO\_Jev} removes all frozen Jev semantic representations ($\boldsymbol{C}_i^{\mathrm{top}}, \boldsymbol{Q}_j^{\mathrm{top}}, \boldsymbol{E}_{ij}^{\mathrm{top}}$, and both evidence-present and prior-free versions of $\boldsymbol{J}_{ij}$) and relies solely on the continuous covariates ($\boldsymbol{x}_i^C, \boldsymbol{x}_j^Q, \boldsymbol{x}_{ij}^E$); (2)~\texttt{NO\_JOINT} removes both evidence-present and prior-free upper-layer cross-stream joint summaries $\boldsymbol{J}_{ij}$ while retaining the independent layer-32 semantic landmarks; (3)~\texttt{FLAT} replaces the hierarchical change--polarity probability factorization with a flat three-class softmax head; (4)~\texttt{NO\_RANK} disables the cross-target pairwise ranking loss ($\mathcal{L} = \mathcal{L}_{\mathrm{main}}$); and (5)~\texttt{NO\_MASK} disables synchronized historical-evidence dropout during training ($p_{\mathrm{ev}} = 0$).
Beyond these retraining-based ablations, \textbf{Target permutation} is evaluated directly on the trained full model by randomly shuffling the perturbation target $i$ across test instances while keeping the query gene $j$ and its historical evidence unchanged.
A predictor that merely recites the historical response propensity of gene $j$ would remain unaffected by target shuffling, so this mismatch directly tests whether S1Pert relies on the actual upstream intervention.

Table~\ref{tab:ablation_summary} identifies semantic encoding and target-conditioned learning as complementary contributors to S1Pert's performance, with the complete model achieving the highest overall Macro-F1 and F1 for all three classes. Jev encoding provides the largest component-level gain: \texttt{NO\_Jev} reduces Macro-F1 from $0.769$ to $0.739$ and minority Downregulated F1 from $0.602$ to $0.542$, demonstrating the value of semantic information beyond numerical covariates. Cross-stream contextualization further strengthens these representations. Without upper-layer joint interaction, \texttt{NO\_JOINT} yields Macro-F1 of $0.755$, Unchanged F1 of $0.923$, and Downregulated F1 of $0.582$, supporting the contribution of the joint summaries $\boldsymbol{J}_{ij}$ to target--query integration.

The decoding and training components reinforce minority-response discrimination. Replacing the hierarchical head with \texttt{FLAT} lowers Downregulated F1 to $0.589$. Removing cross-target ranking or synchronized evidence dropout reduces Macro-F1 to $0.760$ and $0.760$, respectively, with corresponding Downregulated F1 scores of $0.587$ and $0.584$. The importance of target identity is most evident under test-time target permutation: Macro-F1 falls by $6.630$ percentage points to $0.703$, while Downregulated F1 falls by $11.300$ points to $0.489$. This sensitivity shows that S1Pert uses the perturbed target to distinguish response states rather than relying solely on query-gene response propensities. 

In summary, S1Pert combines multilevel System-One semantic encoding with prior-deconfounded hierarchical decoding to improve target-specific response prediction, particularly for the minority downregulated class.

\subsection{Prior-Stress Deconfounding Evaluation}

Aggregate classification scores in Table~\ref{tab:s1pert_main_comparison} can be inflated by memorizing dominant gene-level outcomes~\cite{yuan2026plausibility}. Across all 20 target-held-out splits, we therefore compare S1Pert against an unconditioned historical-propensity baseline (P1, outputting $\boldsymbol{p}_{ij}^{\mathrm{ev}}$) and the top representative from each baseline family: GenePT-Prot, STATE, GEARS, and ALIGNED (Fig.~\ref{fig:prior_stress}). Holding each query gene $j$ fixed across unseen targets, we compute within-gene Change AUC (Chg AUC, changed vs.\ unchanged) and Polarity AUC (Pol AUC, upregulated vs.\ downregulated)~\cite{yuan2026plausibility,wu2025perturbqa}. We further evaluate robustness across prior-entropy tertiles $\mathrm{Ent}(\boldsymbol{p}_{ij}^{\mathrm{ev}})$, counter-propensity reversals ($y_{ij} \neq \arg\max_k p_{ij,k}^{\mathrm{ev}}$), and training-support strata $S_{ij}^q$.
In Fig.~\ref{fig:prior_stress} (a), plotting pooled Macro-F1 against within-gene Pol AUC sharply separates target-specific discrimination from prior memorization. Both P1 ($0.738$ Macro-F1) and ALIGNED ($0.715$ Macro-F1) collapse onto the no-discrimination boundary ($\approx 0.500$ Pol AUC, $\le 0.514$ Chg AUC), revealing that ALIGNED achieves high pooled accuracy primarily by reciting gene-level historical tendencies rather than resolving target-specific regulation. Conversely, GenePT-Prot, STATE, and GEARS achieve moderate ranking ($0.604$--$0.667$ Pol AUC) but low Macro-F1 ($0.613$--$0.663$), whereas S1Pert dominates the upper-right frontier with the highest Macro-F1 ($0.774$) and Pol AUC ($0.697$), alongside a Chg AUC of $0.735$. In Fig.~\ref{fig:prior_stress} (b), S1Pert leads all competitors across prior-entropy tertiles, reaching $0.778$ and $0.641$ Macro-F1 on medium- and high-entropy genes lacking dominant historical tendencies. Under counter-propensity reversals, P1 and ALIGNED plummet to $0.000$ and $0.245$ Down-F1, while S1Pert reaches $0.450$. Across decreasing training support $S_{ij}^q$ in Fig.~\ref{fig:prior_stress} (c), S1Pert ranks first in every regime and retains $0.353$ Macro-F1 at zero-support cold start ($S_{ij}^q = 0$, vs.\ $0.041$ for P1), confirming that Jev semantics and basal control states sustain prediction without historical priors.

\subsection{Statistical Label Fidelity and Boundary Generalization}

Algorithm~1 requires two-one-sided equivalence testing ($q_{ij}^{\mathrm{eq}} \le \alpha$) to assign verified invariance ($y_{ij} = 1$) and isolates underpowered observations failing both directional and equivalence tests as inconclusive ($y_{ij} = \bot$). Since ground-truth transcriptional states are unobservable in real screens, disjoint cell-pool validation is conducted on K562 by splitting the cells of each perturbation target and the untreated control group into two non-overlapping pools. Lower-depth Pool A ($1/3$ of cells) emulates limited single-cell sampling, while higher-depth Pool B ($2/3$ of cells) provides high-power reference outcomes. Algorithm~1 (S1Pert TOST) is first compared against conventional single-test thresholding (Naive DE, where pairs with $q_{ij}^{\mathrm{up}} > \alpha$ and $q_{ij}^{\mathrm{down}} > \alpha$ are labeled unchanged) across perturbed cell-count strata ($n_i$). Trained K562 models are then evaluated without retraining on weak-signal boundary pairs ($y_{ij} = \bot$) that were withheld from training supervision (Fig.~\ref{fig:label_boundary}).

Fig.~\ref{fig:label_boundary} (a) tracks the $76{,}047$ reference DE pairs ($y_{ij} \in \{0, 2\}$) identified in Pool B when re-evaluated in lower-depth Pool A. Reduced cell depth in Pool A leaves only $32.6\%$ of the reference DE responses directly detectable as DE (Pool-A DE). The remaining $67.4\%$ ($75.5\%$--$76.9\%$ at $n_i < 80$ and $57.6\%$ at $n_i \ge 150$) become statistically underpowered (Pool-A unresolved). Treating nonsignificance as invariance under Naive DE mislabels this entire $67.4\%$ underpowered fraction as unchanged ($y_{ij} = 1$) and corrupts both training labels and historical priors $\boldsymbol{p}_{ij}^{\mathrm{ev}}$. By contrast, S1Pert TOST isolates all underpowered pairs as inconclusive ($y_{ij} = \bot$) without introducing false-invariance labels. In Fig.~\ref{fig:label_boundary} (b), S1Pert TOST achieves $>99.99\%$ cross-pool concordance on mutually resolved pairs ($n = 70.3\text{k}$ overall), whereas Naive DE drops to $79.8\%$ overall ($74.9\%$--$83.2\%$ across $n_i$ strata). Isolating $y_{ij} = \bot$ during training also preserves test-time inference on weak-signal pairs. Fig.~\ref{fig:label_boundary} (c) compares the five target-held-out K562 models against P1 on Pool-B-resolved pairs that were underpowered in Pool A (Pool-A Unresolved, mean $55{,}562$ pairs per split) or inconclusive in the combined cell pool (Full-Pool Unknown, mean $2{,}453$ pairs per split). S1Pert surpasses P1 on all five metrics across both subsets, reaching $0.792$ Macro-F1 and $0.648$ Down-F1 ($+11.4$ points over P1) on Pool-A Unresolved alongside $0.651$ Macro-F1 and $0.733$ Down-F1 ($+15.5$ points over P1) on Full-Pool Unknown. Equivalence-based adjudication therefore protects training supervision and historical priors from low-depth sampling noise while enabling reliable generalization to unresolved boundary regimes.

\begin{figure}[!t]
    \centering
    \includegraphics[width=\columnwidth]{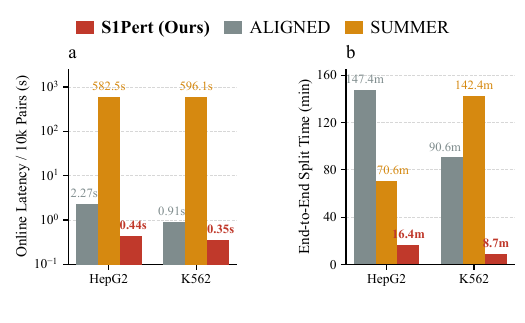}
    \caption{Runtime comparison on HepG2 and K562 using a single NVIDIA RTX PRO 6000 Blackwell (96\,GB) GPU. (a) Online inference latency in seconds per $10{,}000$ test pairs (base-10 logarithmic axis). (b) End-to-end runtime in minutes per split, excluding hyperparameter search. S1Pert's one-time encoding cost is amortized across five splits and added to per-split decoder training and inference.}
    \label{fig:runtime_efficiency}
\end{figure}

\subsection{Runtime Efficiency}
Transcriptome-scale perturbation screening requires accurate prediction with low computational overhead~\cite{barry2024robust,heumos2026pertpy}. Lightweight embedding and regression baselines run quickly, but sacrifice substantial accuracy in Table~\ref{tab:s1pert_main_comparison}. Conversely, the two strongest external competitors (ALIGNED and SUMMER) incur severe computational bottlenecks. Fig.~\ref{fig:runtime_efficiency} compares online latency per $10{,}000$ test pairs and end-to-end runtime per split on HepG2 and K562 essential using one NVIDIA RTX PRO 6000 Blackwell (96\,GB) GPU. Online latency measures test-split inference only, whereas end-to-end runtime per split covers the complete pipeline for one split without hyperparameter search. For S1Pert, one-time 4B encoding is amortized across the five splits and added to per-split decoder training and inference. ALIGNED executes two-round abductive training and test evaluation per split. As a strong baseline, SUMMER has a per-split end-to-end runtime identical to its full-split online Qwen3-8B rationale decoding time. In Fig.~\ref{fig:runtime_efficiency} (a), S1Pert requires only $0.44\,\mathrm{s}$ and $0.35\,\mathrm{s}$ per $10{,}000$ pairs. This online latency is $2.6\times\text{--}5.2\times$ faster than ALIGNED ($2.27\,\mathrm{s}$ and $0.91\,\mathrm{s}$) and $1{,}330\times\text{--}1{,}720\times$ faster than SUMMER ($582.5\,\mathrm{s}$ and $596.1\,\mathrm{s}$). In Fig.~\ref{fig:runtime_efficiency} (b), S1Pert requires $16.4\,\mathrm{min}$ and $8.7\,\mathrm{min}$ per split, running $9.0\times\text{--}10.4\times$ faster than ALIGNED ($147.4\,\mathrm{min}$ and $90.6\,\mathrm{min}$) and $4.3\times\text{--}16.4\times$ faster than SUMMER ($70.6\,\mathrm{min}$ and $142.4\,\mathrm{min}$). Offline landmark caching thus avoids repeated backbone computation during online inference and makes foundation-model perturbation screening practical at transcriptome scale.

\section{Conclusion}
In this work, we propose S1Pert, a unified framework for directional perturbation decision prediction across unseen genetic targets. By integrating equivalence-based statistical adjudication, split-cached foundation-model landmark encoding, and factorized cross-attentive readout, S1Pert addresses key challenges in filtering low-depth sampling noise, overcoming foundation-model inference bottlenecks, and preventing historical prior shortcuts. Experiments on four real-world single-cell perturbation benchmarks show that S1Pert outperforms existing baselines in directional decision accuracy, within-gene target ranking, and runtime efficiency. Designed for transcriptome-scale screening, S1Pert serves as a practical \textit{in silico} prioritization tool that narrows down candidate genetic interventions before wet-lab validation. Future work will analyze boundary failure cases, extend the framework to combinatorial multi-gene perturbations and multimodal cellular signals, and scale to broader tissue contexts while preserving statistical rigor and efficiency.
                         
\section*{AI-Generated Content Acknowledgment}
OpenAI Codex was used to assist with code development and debugging for the methods and experiments in Sections~III and~IV, and with manuscript drafting and language refinement in Sections~I--IV. The authors take full responsibility for the manuscript and code.

\bibliographystyle{IEEEtran}
\bibliography{cite}

@article{dixit2016perturb,
  author    = {Dixit, Atray and Parnas, Oren and Li, Biyu and Chen, Jenny and Fulco, Charles P. and Jerby-Arnon, Livnat and Marjanovic, Nemanja D. and Dionne, Danielle and Burks, Tyler and Raychowdhury, Raktima and Adamson, Britt and Norman, Thomas M. and Lander, Eric S. and Weissman, Jonathan S. and Friedman, Nir and Regev, Aviv},
  title     = {Perturb-seq: Dissecting Molecular Circuits with Scalable Single-Cell {RNA} Profiling of Pooled Genetic Screens},
  journal   = {Cell},
  volume    = {167},
  number    = {7},
  pages     = {1853--1866.e17},
  year      = {2016},
  publisher = {Elsevier},
  doi       = {10.1016/j.cell.2016.11.038}
}

@article{replogle2022mapping,
  author    = {Replogle, Joseph M. and Saunders, Reuben A. and Pogson, Angela N. and Hussmann, Jeffrey A. and Lenail, Alexander and Guna, Alina and Mascibroda, Lauren and Wagner, Eric J. and Adelman, Karen and Lithwick-Yanai, Gila and Iremadze, Nika and Oberstrass, Florian and Lipson, Doron and Bonnar, Jessica L. and Jost, Marco and Norman, Thomas M. and Weissman, Jonathan S.},
  title     = {Mapping Information-Rich Genotype-Phenotype Landscapes with Genome-Scale {Perturb-seq}},
  journal   = {Cell},
  volume    = {185},
  number    = {14},
  pages     = {2559--2575.e28},
  year      = {2022},
  publisher = {Elsevier},
  doi       = {10.1016/j.cell.2022.05.013}
}

@article{cui2024scgpt,
  author    = {Cui, Haotian and Wang, Chloe and Maan, Hassaan and Pang, Kuan and Luo, Fengning and Duan, Nan and Wang, Bo},
  title     = {{scGPT}: Toward Building a Foundation Model for Single-Cell Multi-Omics Using Generative {AI}},
  journal   = {Nature Methods},
  volume    = {21},
  number    = {8},
  pages     = {1470--1480},
  year      = {2024},
  publisher = {Nature Publishing Group},
  doi       = {10.1038/s41592-024-02201-0}
}

@inproceedings{xiang2026aligned,
  author    = {Xiang, Yuanfang and Ai, Lun},
  title     = {Adaptive Data-Knowledge Alignment in Genetic Perturbation Prediction},
  booktitle = {Proceedings of the 14th International Conference on Learning Representations (ICLR)},
  year      = {2026}
}

@article{phillips2025synthpert,
  author    = {Phillips, Lawrence and Martell, Marc Boubnovski and Misra, Aditya and Stoisser, Josefa Lia and Prada-Medina, Cesar A. and Donovan-Maiye, Rory and M{\"a}rtens, Kaspar},
  title     = {{SynthPert}: Enhancing {LLM} Biological Reasoning via Synthetic Reasoning Traces for Cellular Perturbation Prediction},
  journal   = {arXiv preprint arXiv:2509.25346},
  year      = {2025}
}

@article{love2014moderated,
  author    = {Love, Michael I. and Huber, Wolfgang and Anders, Simon},
  title     = {Moderated Estimation of Fold Change and Dispersion for {RNA-seq} Data with {DESeq2}},
  journal   = {Genome Biology},
  volume    = {15},
  number    = {12},
  pages     = {550},
  year      = {2014},
  publisher = {BioMed Central},
  doi       = {10.1186/s13059-014-0550-8}
}

@article{wenkel2026txpert,
  author    = {Wenkel, Frederik and Tu, Wilson and Masschelein, Cassandra and Shirzad, Hamed and Hodgson, Liam and Bendidi, Ihab and Eastwood, Cian and Whitfield, Shawn T. and Russell, Craig and El Mesbahi, Yassir and Ding, Jiarui and Fay, Marta M. and Earnshaw, Berton and Noutahi, Emmanuel and Denton, Alisandra K.},
  title     = {{TxPert}: Using Multiple Knowledge Graphs for Prediction of Transcriptomic Perturbation Effects},
  journal   = {Nature Biotechnology},
  year      = {2026},
  publisher = {Nature Publishing Group},
  doi       = {10.1038/s41587-026-03113-4}
}

@inproceedings{wang2026aroma,
  author    = {Wang, Zhenyu and Ye, Geyan and Liu, Wei and Ng, Man Tat Alexander},
  title     = {{AROMA}: Augmented Reasoning Over a Multimodal Architecture for Virtual Cell Genetic Perturbation Modeling},
  booktitle = {Findings of the Association for Computational Linguistics: ACL 2026},
  pages     = {31855--31881},
  year      = {2026},
  month     = jul,
  publisher = {Association for Computational Linguistics},
  doi       = {10.18653/v1/2026.findings-acl.1594}
}

@article{kim2026pertreason,
  author    = {Kim, Dongkwan and Gao, Yiming and Yang, Yining and Shen, Yang},
  title     = {{PertReason}: A Knowledge-Grounded Benchmark and Framework for Cell-State-Conditioned Mechanistic Reasoning of Perturbation Effects},
  journal   = {arXiv preprint arXiv:2607.18777},
  year      = {2026}
}

@misc{kim2026progressive,
  author        = {Kim, Hyomin and Hwang, Sang-Yeon and Lim, Jaechang and Piao, Yinhua and Oh, Yunhak and Kim, Woo Youn and Park, Chanyoung and Ahn, Sungsoo and Jeon, Junhyeok},
  title         = {Progressive Multi-Agent Reasoning for Biological Perturbation Prediction},
  year          = {2026},
  eprint        = {2602.07408},
  archivePrefix = {arXiv},
  primaryClass  = {cs.AI},
  doi           = {10.48550/arXiv.2602.07408},
  url           = {https://arxiv.org/abs/2602.07408}
}

@article{yuan2026plausibility,
  author    = {Yuan, Xinyu and Liu, Xixian and Zhao, Jianan and Zhang, Yashi and Guo, Hongyu and Tang, Jian},
  title     = {Plausibility Is Not Prediction: Benchmarking Large Language Models for Virtual Cell Perturbation Modeling},
  journal   = {arXiv preprint arXiv:2606.01042},
  year      = {2026}
}

@article{roohani2024gears,
  author    = {Roohani, Yusuf and Huang, Kexin and Leskovec, Jure},
  title     = {Predicting Transcriptional Outcomes of Novel Multigene Perturbations with {GEARS}},
  journal   = {Nature Biotechnology},
  volume    = {42},
  number    = {6},
  pages     = {927--935},
  year      = {2024},
  publisher = {Nature Publishing Group},
  doi       = {10.1038/s41587-023-01905-6}
}

@article{zhu2025scouter,
  author    = {Zhu, Ouyang and Li, Jun},
  title     = {{Scouter} predicts transcriptional responses to genetic perturbations with large language model embeddings},
  journal   = {Nature Computational Science},
  volume    = {6},
  number    = {1},
  pages     = {21--28},
  year      = {2026},
  publisher = {Nature Publishing Group},
  doi       = {10.1038/s43588-025-00912-8}
}

@article{chen2025genept,
  author  = {Chen, Yiqun and Zou, James},
  title   = {Simple and effective embedding model for single-cell biology built from {ChatGPT}},
  journal = {Nature Biomedical Engineering},
  volume  = {9},
  number  = {4},
  pages   = {483--493},
  year    = {2025},
  doi     = {10.1038/s41551-024-01284-6}
}

@misc{mapika2026decider4b,
  author       = {{Mapika}},
  title        = {{DECIDER-4B}: An Open System-One Decision Model for Single-Pass Option Probability Estimation},
  year         = {2026},
  howpublished = {Hugging Face model repository},
  url          = {https://huggingface.co/Mapika/decider-4b},
  note         = {Experimental checkpoint revision:
                  c23ab4d2483e7c6a93484463e9afe1688b29bedd}
}

@article{cai2025genetic,
  author={Cai, Luyang and Yu, Jixang and Lin, Qiuzhen and Liu, Lichen and Li, Xiangtao and Wong, Ka-Chun},
  journal={IEEE Journal of Biomedical and Health Informatics}, 
  title={Genetic Perturbation Modeling for Human Cell Therapy With BRNET}, 
  year={2026},
  volume={30},
  number={5},
  pages={3971-3980}
}

@inproceedings{yang2025earn,
  title={Earn: Efficient inference acceleration for llm-based generative recommendation by register tokens},
  author={Yang, Chaoqun and Lin, Xinyu and Wang, Wenjie and Li, Yongqi and Sun, Teng and Han, Xianjing and Chua, Tat-Seng},
  booktitle={Proceedings of the 31st ACM SIGKDD Conference on Knowledge Discovery and Data Mining V. 2},
  pages={3483--3494},
  year={2025}
}

@article{smirnova2022nessy,
   author={Smirnova, Alisa and Yang, Jie and Yang, Dingqi and Cudre-Mauroux, Philippe},
  journal={IEEE Transactions on Knowledge and Data Engineering}, 
  title={Nessy: A Neuro-Symbolic System for Label Noise Reduction}, 
  year={2023},
  volume={35},
  number={8},
  pages={8300-8311},
}

@article{welch1947generalization,
  title={The generalization of ‘STUDENT'S’problem when several different population varlances are involved},
  author={Welch, Bernard L},
  journal={Biometrika},
  volume={34},
  number={1-2},
  pages={28--35},
  year={1947},
  publisher={Oxford University Press}
}

@article{yang2023learning,
  author={Yang, Hao and Jin, You-Zhi and Li, Zi-Yin and Wang, Deng-Bao and Geng, Xin and Zhang, Min-Ling},
  journal={IEEE Transactions on Knowledge and Data Engineering}, 
  title={Learning From Noisy Labels via Dynamic Loss Thresholding}, 
  year={2024},
  volume={36},
  number={11},
  pages={6503-6516},
  doi={10.1109/TKDE.2023.3313604}}

@inproceedings{chen2026bond,
  title={BOND: A Co-Designed Framework for LLM-Powered Analytics Over Relational Data},
  author={Chen, Lixiang and Zheng, Qin and Pan, Zhicheng and Yang, Chengcheng and Zhang, Rong and Zhou, Xuan},
  booktitle={2026 IEEE 42nd International Conference on Data Engineering (ICDE)},
  pages={1914--1926},
  year={2026},
  organization={IEEE}
}

@inproceedings{du2026c2kv,
  title={C2kv: Compressed and composable kv cache reuse for efficient llm inference},
  author={Du, Chuheng and Chen, Junyi and Tang, Hanlin and Liu, Kan and Lan, Tao and Qu, Lin and Niu, Chaoyue and Liu, Shengzhong and Chen, Guihai and Wu, Fan},
  booktitle={Proceedings of the 32nd ACM SIGKDD Conference on Knowledge Discovery and Data Mining V. 2},
  pages={991--1002},
  year={2026}
}

@inproceedings{shin2026matkv,
  title={MatKV: Trading compute for flash storage in LLM inference},
  author={Shin, Kun-Woo and Park, Jay H and Oh, Moonwook and Jo, Yohan and Do, Jaeyoung and Lee, Sang-Won},
  booktitle={2026 IEEE 42nd International Conference on Data Engineering (ICDE)},
  pages={877--890},
  year={2026},
  organization={IEEE}
}

@inproceedings{li2021cleanml,
  title={Cleanml: A study for evaluating the impact of data cleaning on ml classification tasks},
  author={Li, Peng and Rao, Xi and Blase, Jennifer and Zhang, Yue and Chu, Xu and Zhang, Ce},
  booktitle={2021 IEEE 37th international conference on data engineering (ICDE)},
  pages={13--24},
  year={2021},
  organization={IEEE}
}

@article{satterthwaite1946approximate,
  title={An approximate distribution of estimates of variance components},
  author={Satterthwaite, Franklin E},
  journal={Biometrics bulletin},
  volume={2},
  number={6},
  pages={110--114},
  year={1946},
  publisher={JSTOR}
}

@ARTICLE{11603313,
  author={Cai, Luyang and Yu, Jixiang and Liu, Licheng and Lin, Qiuzhen and Li, Xiangtao and Wong, Ka-Chun},
  journal={IEEE Transactions on Big Data}, 
  title={Context-Aware Dual-Graph Neural Network for Cellular State Inference Under Genetic Perturbations}, 
  year={2026},
  volume={12},
  number={5},
  pages={1938-1949},
  doi={10.1109/TBDATA.2026.3712210}}

@article{barry2024robust,
  title={Robust differential expression testing for single-cell CRISPR screens at low multiplicity of infection},
  author={Barry, Timothy and Mason, Kaishu and Roeder, Kathryn and Katsevich, Eugene},
  journal={Genome biology},
  volume={25},
  number={1},
  pages={124},
  year={2024},
  publisher={Springer}
}

@inproceedings{jiang2026learning,
  title={Learning latent dynamical causal processes for single-cell perturbation prediction},
  author={Jiang, Wenkang and Liu, Yuhang and Gao, Erdun and Abbasnejad, Ehsan and Yao, Lina and Shi, Javen Qinfeng},
  booktitle={Proceedings of the 32nd ACM SIGKDD Conference on Knowledge Discovery and Data Mining V. 2},
  pages={11177--11188},
  year={2026}
}

@article{lakens2017equivalence,
  title={Equivalence tests: A practical primer for t tests, correlations, and meta-analyses},
  author={Lakens, Dani{\"e}l},
  journal={Social psychological and personality science},
  volume={8},
  number={4},
  pages={355--362},
  year={2017},
  publisher={Sage Publications, Sage CA: Los Angeles, CA}
}

@inproceedings{wei2026vcworld,
  title={VCWorld: A biological world model for virtual cell simulation},
  author={Wei, Zhijian and Ma, Runze and Wang, Zichen and Li, Zhongmin and Song, Shuotong and Zheng, Shuangjia},
  booktitle={International Conference on Learning Representations},
  volume={2026},
  pages={72795--72823},
  year={2026}
}

@article{heumos2026pertpy,
  author  = {Heumos, Lukas and Ji, Yuge and May, Lilly and others},
  title   = {Pertpy: An end-to-end framework for perturbation analysis},
  journal = {Nature Methods},
  volume  = {23},
  number  = {2},
  pages   = {350--359},
  year    = {2026},
  doi     = {10.1038/s41592-025-02909-7}
}

@misc{chen2024geneptembeddings,
  author       = {Chen, Yiqun},
  title        = {Gene embeddings used in {GenePT}: A Simple But Hard-to-Beat Foundation Model for Genes and Cells Built From {ChatGPT}},
  year         = {2024},
  howpublished = {Zenodo},
  version      = {v2},
  doi          = {10.5281/zenodo.10833191},
  url          = {https://doi.org/10.5281/zenodo.10833191},
  note         = {Dataset, CC BY 4.0}
}

@misc{qwen3technicalreport,
  title         = {{Qwen3} Technical Report},
  author        = {{Qwen Team}},
  year          = {2025},
  eprint        = {2505.09388},
  archivePrefix = {arXiv},
  primaryClass  = {cs.CL},
  url           = {https://arxiv.org/abs/2505.09388}
}

@article{State2026,
  author  = {Adduri, Abhinav K. and Gautam, Dhruv
             and Bevilacqua, Beatrice and others},
  title   = {Predicting Cellular Responses to Perturbation
             Across Diverse Contexts with {State}},
  journal = {Cell},
  volume  = {189},
  number  = {19},
  pages   = {5914--5931.e20},
  year    = {2026},
  doi     = {10.1016/j.cell.2026.07.052}
}

@article{wu2025perturbqa,
  author    = {Wu, Menghua and Littman, Russell and Levine, Jacob and Qiu, Lin and Biancalani, Tommaso and Richmond, David and H{\"u}tter, Jan-Christian},
  title     = {Contextualizing Biological Perturbation Experiments Through Language},
  journal   = {arXiv preprint arXiv:2502.21290},
  volume    = {2025},
  pages     = {73598--73624},
  year      = {2025}
}

@article{ahlmann2025deep,
  author    = {Ahlmann-Eltze, Constantin and Huber, Wolfgang and Anders, Simon},
  title     = {Deep-Learning-Based Gene Perturbation Effect Prediction Does Not Yet Outperform Simple Linear Baselines},
  volume    = {22},
  number    = {8},
  pages     = {1657--1661},
  journal   = {Nature Methods},
  year      = {2025},
  publisher = {Nature Publishing Group},
  doi       = {10.1038/s41592-025-02772-6}
}

@article{nadig2025transcriptome,
  author    = {Nadig, Ajay and Replogle, Joseph M. and Pogson, Angela N. and McCarroll, Steven A. and Weissman, Jonathan S. and Robinson, Elise B. and O'Connor, Luke J.},
  title     = {Transcriptome-Wide Analysis of Differential Expression in Perturbation Atlases},
  journal   = {Nature Genetics},
  volume    = {57},
  pages     = {1228--1237},
  year      = {2025},
  publisher = {Nature Publishing Group}
}

@article{schuirmann1987comparison,
  author    = {Schuirmann, Donald J.},
  title     = {A Comparison of the Two One-Sided Tests Procedure and the Power Approach for Assessing the Equivalence of Average Bioavailability},
  journal   = {Journal of Pharmacokinetics and Biopharmaceutics},
  volume    = {15},
  number    = {6},
  pages     = {657--680},
  year      = {1987},
  publisher = {Springer},
  doi       = {10.1007/BF01068419}
}

@article{krishnan2016activeclean,
  author  = {Krishnan, Sanjay and Wang, Jiannan and Wu, Eugene
             and Franklin, Michael J. and Goldberg, Ken},
  title   = {{ActiveClean}: Interactive Data Cleaning For Statistical Modeling},
  journal = {Proceedings of the VLDB Endowment},
  volume  = {9},
  number  = {12},
  pages   = {948--959},
  year    = {2016},
  doi     = {10.14778/2994509.2994514}
}

@article{rekatsinas2017holoclean,
  author  = {Rekatsinas, Theodoros and Chu, Xu and Ilyas, Ihab F.
             and R{\'e}, Christopher},
  title   = {{HoloClean}: Holistic Data Repairs with Probabilistic Inference},
  journal = {Proceedings of the VLDB Endowment},
  volume  = {10},
  number  = {11},
  pages   = {1190--1201},
  year    = {2017},
  doi     = {10.14778/3137628.3137631}
}

@article{patel2025lotus,
  author  = {Patel, Liana and Jha, Siddharth and Pan, Melissa
             and Gupta, Harshit and Asawa, Parth
             and Guestrin, Carlos and Zaharia, Matei},
  title   = {Semantic Operators and Their Optimization: Enabling
             {LLM}-Based Data Processing with Accuracy Guarantees in {LOTUS}},
  journal = {Proceedings of the VLDB Endowment},
  volume  = {18},
  number  = {11},
  pages   = {4171--4184},
  year    = {2025},
  doi     = {10.14778/3749646.3749685}
}

@article{rood2024toward,
  title={Toward a foundation model of causal cell and tissue biology with a Perturbation Cell and Tissue Atlas},
  author={Rood, Jennifer E and Hupalowska, Anna and Regev, Aviv},
  journal={Cell},
  volume={187},
  number={17},
  pages={4520--4545},
  year={2024},
  publisher={Elsevier}
}

@article{peidli2024scperturb,
  title={scPerturb: harmonized single-cell perturbation data},
  author={Peidli, Stefan and Green, Tessa D and Shen, Ciyue and Gross, Torsten and Min, Joseph and Garda, Samuele and Yuan, Bo and Schumacher, Linus J and Taylor-King, Jake P and Marks, Debora S and Luna, Augustin and Bl{\"u}thgen, Nils and Sander, Chris},
  journal={Nature Methods},
  volume={21},
  number={3},
  pages={531--540},
  year={2024}
}

@inproceedings{wei2021model,
  title={Model-agnostic counterfactual reasoning for eliminating popularity bias in recommender system},
  author={Wei, Tianxin and Feng, Fuli and Chen, Jiawei and Wu, Ziwei and Yi, Jinfeng and He, Xiangnan},
  booktitle={Proceedings of the 27th ACM SIGKDD Conference on Knowledge Discovery \& Data Mining (KDD)},
  pages={1791--1800},
  year={2021}
}

@inproceedings{tan2025taming,
  title={Taming recommendation bias with causal intervention on evolving personal popularity},
  author={Tan, Shiyin and Li, Dongyuan and Jiang, Renhe and Wang, Zhen and Yu, Xingtong and Okumura, Manabu},
  booktitle={Proceedings of the 31st ACM SIGKDD Conference on Knowledge Discovery and Data Mining (KDD)},
  pages={2790--2800},
  year={2025}
}

@inproceedings{liu2025palimpzest,
  title={Palimpzest: Optimizing AI-powered analytics with declarative query processing},
  author={Liu, Chunwei and Russo, Matthew and Cafarella, Michael and Cao, Lei and Chen, Peter Baile and Chen, Zui and Franklin, Michael and Kraska, Tim and Madden, Samuel and Shahout, Rana and Vitagliano, Gerardo},
  booktitle={Proceedings of the 15th Conference on Innovative Data Systems Research (CIDR)},
  year={2025}
}

@inproceedings{gim2024prompt,
  title={Prompt Cache: Modular attention reuse for low-latency inference},
  author={Gim, In and Chen, Guojun and Lee, Seung-seob and Sarda, Nikhil and Khandelwal, Anurag and Zhong, Lin},
  booktitle={Proceedings of Machine Learning and Systems (MLSys)},
  volume={6},
  pages={325--338},
  year={2024}
}

\end{document}